\documentclass[11pt,a4paper]{article}

\usepackage[truedimen,margin=30mm]{geometry} 

\usepackage{mathrsfs}
\usepackage{amssymb}
\usepackage{amsmath}
\usepackage{ascmac}
\usepackage{amsthm}

\usepackage[pdftex]{graphicx}
\usepackage[pdftex]{color}

\usepackage{color}
\usepackage{natbib}
\usepackage{setspace}
\usepackage{url}

\usepackage{titlesec}
\titleformat*{\section}{\large\bfseries}
\titleformat*{\subsection}{\it}

\def\by{{\text{\boldmath$y$}}}
\def\bx{{\text{\boldmath$x$}}}
\def\bPsi{{\text{\boldmath$\Psi$}}}
\def\bpsi{{\text{\boldmath$\psi$}}}
\def\bomega{{\text{\boldmath$\omega$}}}
\def\bth{{\text{\boldmath$\theta$}}}
\def\bal{{\text{\boldmath$\alpha$}}}
\def\bbe{{\text{\boldmath$\beta$}}}
\def\bEta{{\text{\boldmath$\eta$}}}
\def\bpi{{\text{\boldmath$\pi$}}}
\def\bmu{{\text{\boldmath$\mu$}}}
\def\bSig{{\text{\boldmath$\Sigma$}}}
\def\bOmega{{\text{\boldmath$\Omega$}}}

\def\ep{{\varepsilon}}

\def\gam{\gamma}
\def\Re{\mathcal{R}}

\title{{\bf Robust Fitting of Mixture Models using Weighted Complete Estimating Equations}}
\date{}

\begin{document}

\maketitle
\doublespacing

\vspace{-1.5cm}
\begin{center}
{\large Shonosuke Sugasawa$^1$ and Genya Kobayashi$^2$ }
\end{center}

\medskip
\noindent
$^1$Center for Spatial Information Science, The University of Tokyo\\
$^2$Graduate School of Social Science, Chiba University

\vspace{5mm}
\begin{center}
{\bf \large Abstract}
\end{center}
Mixture modeling, which considers the potential heterogeneity in data, is widely adopted for classification and clustering problems. Mixture models can be estimated using the Expectation-Maximization algorithm, which works with the complete estimating equations conditioned by the latent membership variables of the cluster assignment based on the hierarchical expression of mixture models. However, when the mixture components have light tails such as a normal distribution, the mixture model can be sensitive to outliers. This study proposes a method of weighted complete estimating equations (WCE) for the robust fitting of mixture models. Our WCE introduces weights to complete estimating equations such that the weights can automatically downweight the outliers. The weights are constructed similarly to the density power divergence for mixture models, but in our WCE, they depend only on the component distributions and not on the whole mixture. A novel expectation-estimating-equation (EEE) algorithm is also developed to solve the WCE. For illustrative purposes, a multivariate Gaussian mixture, a mixture of experts, and a multivariate skew normal mixture are considered, and how our EEE algorithm can be implemented for these specific models is described. The numerical performance of the proposed robust estimation method was examined using simulated and real datasets.

\vspace{-0cm}


\newpage
\section{Introduction}\label{sec:intro}

Mixture modeling \citep{Mc2004} is very popular for distribution estimation, regression, and model-based clustering by taking account of potential subgroup structures of data.
Typically, such mixture models are fitted by the maximum likelihood method using the well-known Expectation-Maximization (EM) algorithm.
However, data often contain outliers and the maximum likelihood method can be highly affected by them. 
The presence of outliers would result in the biased and inefficient statistical inference of the parameters of interest and would thus make recovering the underlying clustering structure of the data very difficult. 
A typical remedy for this problem is to replace the normal component distributions with heavy-tailed components. 
For example, \cite{bagnato2017multivariate}, \cite{forbes2014new}, \cite{PM2000}, \cite{punzo2021multiple}, \cite{sun2010robust}, and \cite{zhang2010robust} used symmetric component distributions, \cite{basso2010robust}, \cite{lee2014finite}, \cite{lin2010robust}, \cite{Morris2019}, and \cite{wang2009multivariate} used skewed distributions, and \cite{galimberti2014multivariate}, \cite{ingrassia2014model}, \cite{mazza2020mixtures}, \cite{punzo2017robust}, \cite{Song2014}, \cite{yao2014robust}, and \cite{zarei2019use} considered robustifying the mixture of regressions. 
However, this approach cannot distinguish non-outliers from outliers straightforwardly because it fits a mixture model to all the observations, including outliers. 
Apart from using heavy-tailed distributions, a class of generalized likelihood is an effective alternative tool for the robust fitting of statistical models. 
In the context of mixture modeling, \cite{Fujisawa2006} employed density power divergence \citep{Basu1998} for robust estimation of Gaussian mixture models.  
However, a direct application of the divergence may suffer from computational problems because the objective functions may contain intractable integrals.
For example, the objective function under the density power divergence includes the integral $\int f(x;\theta)^{\gamma}dx$, where $f(x;\theta)$ is the density function with the parameter $\theta$ and $\gamma$ is the tuning parameter. 
When $f(x;\theta)$ is the density function of a mixture model, the integral becomes intractable. 
Therefore, these approaches are not necessarily appealing in practice, despite their desirable robustness properties.

In this paper, we develop a new approach to robust mixture modeling by newly introducing the idea of the weighted complete estimating equations (WCE). 
Complete estimating equations appear in the EM algorithm and are conditioned on the latent membership variables of the cluster assignment. 
Our WCE introduces weights to the complete estimating equations such that the contribution of outliers to the complete estimating equations is automatically downweighted. 
The weight is defined based on the assumed component distributions. 
By conditioning the membership variables in WCE, only the weighted estimating equations for each component distribution must be considered separately instead of directly using the weighted estimating equations for a mixture model. 
Since the derived WCE depends on the unknown latent membership variables, these latent variables are augmented via their posterior expectations to solve the WCE, which provides an expectation-estimating-equation (EEE) algorithm \citep{EEE}. 
The proposed EEE algorithm is general and can be applied to various mixture models.
The proposed WCE method is then applied to three types of mixture models: a multivariate Gaussian mixture, a mixture of experts \citep{Jacob1991}, and a multivariate skew-normal mixture \citep{Lin2007,Lin2009}. 
For the multivariate Gaussian mixture, the updating steps of the proposed EEE algorithm are obtained in closed form without requiring either numerical integration or optimization steps.
A similar algorithm can be derived for a mixture of experts when the component distributions are Gaussian.
Moreover, for the multivariate skew normal mixture, by introducing additional latent variables based on the stochastic representation of the skew normal distribution, a novel EEE algorithm can be obtained as a slight extension of the proposed general EEE algorithm in which all the updating steps proceed analytically.

The proposed framework of the weighted complete estimating equations unifies several existing approaches regarding the choice of the weight function, such as hard trimming \citep[e.g.][]{tclust}, soft trimming \citep[e.g.][]{Campbell1984,FG2015,Greco2020} and their hybrid \citep[e.g.][]{farcomeni2020robust}.
Among several designs for the weight function, this study adopts the density weight, which leads to theoretically valid and computationally tractable robust estimation procedures not only for the Gaussian mixture but also for a variety of mixture models.

As a related work, \cite{Greco2020} employed a similar idea using the weighted likelihood for the robust fitting of the multivariate Gaussian mixture. 
They constructed weights by first obtaining a kernel density estimate for the Mahalanobis distance concerning the component parameters and then computing the Pearson residuals. 
Since their method requires knowledge of the distribution of the quadratic form of the residuals, their approach cannot be directly extended to other mixture models.
On the other hand, the proposed method can be applied to various mixture models, as in the proposed weight design, and the weights are automatically determined by the component-wise density.
It should be noted that introducing the density weight in estimating equations is closely related to the density power divergence \citep{Basu1998} in which the density power divergence induces the density-weighted likelihood equations. 
However, the novelty of the proposed approach is that the density weight is considered within the framework of complete estimating equations rather than likelihood equations, which leads to tractable robust estimation algorithms for a variety of mixture models.

The remainder of this paper is organized as follows.
In Section~\ref{sec:WEE}, we first describe the proposed WCE method for a general mixture model and derive a general EEE algorithm for solving the WCE.
Then, the general algorithm is applied to the specific mixture models, a multivariate Gaussian mixture (Section~\ref{sec:GM}), a mixture of experts (Section~\ref{sec:MOE}), and a multivariate skew normal mixture (Section~\ref{sec:SNM}). 
In Section~\ref{sec:sim}, the proposed method is demonstrated for a multivariate Gaussian mixture and a skew normal mixture through the simulation studies. 
Section~\ref{sec:app} illustrates practical advantages of the proposed method using real data.
Finally, conclusions and discussions are presented in Section~\ref{sec:dis}.
The R code for implementing the proposed method is available at the GitHub repository (\url{https://github.com/sshonosuke/RobMixture}).

\section{Weighted Estimating Equations for Mixture Modeling}\label{sec:WEE}

\subsection{Weighted complete estimating equations}
Let $\by_1,\ldots,\by_n$ be the random variables on $\mathbb{R}^p$. 
We consider the following mixture model with probability density function $f_M$:
\begin{equation}\label{Mix}
f_M(\by_i;\bPsi)=\sum_{k=1}^K\pi_k f(\by_i;\bth_k),\quad i=1,\dots,n,
\end{equation}
where $\bth_k$ is the set of model parameters in the $k$th component, $\bpi=(\pi_1,\ldots,\pi_K)$ is the vector of grouping or prior membership probabilities, and $\bPsi=(\bpi,\bth_1,\ldots,\bth_K)$ is the collection of all model parameters. 
To fit model \eqref{Mix}, we introduce the latent membership variable $z_i$ defined as $P(z_i=k)=\pi_k$ for $i=1,\dots,n$; therefore, the conditional distribution of $\by_i$ given $z_i=k$ is $f(\by_i;\theta_k)$. 
For notional simplicity, we define $u_{ik}=I(z_i=k)$, which is the indicator function for the condition $z_i=k$.
The complete estimating equations for $\bPsi$ given $z_i$ are given as follows:
\begin{align*}
&\sum_{i=1}^nu_{ik}\frac{\partial}{\partial\bth_k}\log f(\by_i;\bth_k)=0,\\
&\sum_{i=1}^n\frac{u_{ik}}{\pi_k}-\sum_{i=1}^n\frac{u_{iK}}{\pi_K}=0, 
\end{align*}
for $k=1,\dots, K$.

To robustify the estimating equations against outliers, we introduce the component-specific weight $w(\by_i;\bth_k)$ for $i=1,\ldots,n$ and $k=1,\ldots,K$ which controls the contribution from the $i$th observation.
We then propose the following weighted complete estimating equations (WCE): 
\begin{equation}\label{WEE}
\begin{split}
&\sum_{i=1}^nu_{ik}\left\{w(\by_i;\bth_k)\frac{\partial}{\partial\bth_k}\log f(\by_i;\bth_k)-C(\bth_k)\right\}=0,\\
&\sum_{i=1}^n\frac{u_{ik}}{\pi_k}\frac{w(\by_i;\bth_k)}{B(\bth_k)}-\sum_{i=1}^n\frac{u_{iK}}{\pi_K}\frac{w(\by_i;\bth_K)}{B(\bth_K)}=0,
\end{split}
\end{equation}
for $k=1,\dots,K$, where $w(\by_i;\bth_k)$ is the weight function which may depend on the tuning parameter $\gam$ and 
\begin{align*}
&C(\bth_k)
=\int_{\mathbb{R}^p}f(t;\bth_k)w(t;\bth_k)\frac{\partial}{\partial\bth_k}\log f(t;\bth_k)dt,\\
& \ \ \ \ 
B(\bth_k)
=\int_{\mathbb{R}^p}f(t;\bth_k)w(t;\bth_k)dt.
\end{align*}
Note that $B(\bth_k)$ and $C(\bth_k)$ are necessary to make  WCE \eqref{WEE} unbiased. 
That is, the expectations of the estimating equations in \eqref{WEE} with respect to the joint distribution of $\by_i$ and $z_i$ are zero; otherwise, asymptotic properties such as consistency and asymptotic normality of the resulting estimator may not be guaranteed. 
In the unbiased WCE, we consider the specific form of the weight function based on the component density given by 
$$
w(\by_i;\bth_k)=f(\by_i;\bth_k)^{\gam}
$$ 
for $\gam>0$.
The weight is small if $\by_i$ is an outlier, that is, $\by_i$ is located far in the tail of the distribution of the $k$th component $f(\by_i;\bth_k)$.
The weighted estimating equations are reduced to the original complete estimating equation when $\gamma=0$. 
It should be noted that the proposed weight design is not a direct application of the density power divergence of \cite{Basu1998} to the mixture model \eqref{Mix}, which would have used the whole mixture density \eqref{Mix} as the weight in the estimating equations.

\subsection{EEE algorithm}
Since the proposed WCE \eqref{WEE} includes the latent variable $u_{ik}$, WCE should be imputed with the conditional expectation of $u_{ik}$ given $\by_i$.
Starting from some initial values, we propose an EEE algorithm that updates the estimates in the $s$th iteration as follows:

\begin{itemize}

\item[-]
{\it E-step:} \ \ 
Compute the posterior probability:
\begin{equation}\label{post.prob}
u_{ik}^{(s)}=\frac{\pi_k^{(s)}f(\by_i;\bth_k^{(s)})}{\sum_{\ell=1}^K\pi_\ell^{(s)}f(\by_i;\bth_\ell^{(s)})}, \ \ \ \ k=1,\ldots,K.
\end{equation}

\item[-]
{\it EE-step:} \ \ 
Update the membership probabilities $\pi_k$'s as
\begin{equation}\label{pi-update}
\pi_k^{(s+1)}=\frac{\sum_{i=1}^nu_{ik}^{(s)}w_{ik}^{(s)}/B(\bth_k^{(s)})}{\sum_{\ell=1}^K\sum_{i=1}^nu_{i\ell}^{(s)}w_{ik}^{(s)}/B(\bth_\ell^{(s)})},
\end{equation}
and the component-specific parameters $\bth_k$ by solving the estimating equations:
$$
\sum_{i=1}^nu_{ik}^{(s)}\left\{w_{ik}^{(s)}\frac{\partial}{\partial\bth_k}\log f(\by_i;\bth_k)-C(\bth_k)\right\}=0,
$$
where $w_{ik}^{(s)}=w(\by_i;\bth_k^{(s)})$.
\end{itemize}

Note that the updating process \eqref{pi-update} can be obtained by solving the imputed version of the second equation in \eqref{WEE}.
To apply the general algorithm above to a particular mixture model such as a multivariate Gaussian mixture, the only requirement is to calculate the bias correction terms $C(\bth_k)$ and $B(\bth_k)$.
As shown in Section \ref{sec:GM}, the bias correction terms are quite simple under a Gaussian distribution.
Moreover, the algorithm can be easily modified to a case in which the distribution of each component admits a hierarchical or stochastic representation. 
For instance, the multivariate skew normal distribution has a hierarchical representation based on the multivariate normal distribution, which allows us to derive tractable weighted complete estimating equations to carry out the proposed robust EEE algorithm, as demonstrated in Section \ref{sec:SNM}.

Regarding the setting of the starting values, we can use the estimation results of some existing robust methods. 
The detailed initialization strategy for each mixture model is discussed in Sections~\ref{sec:GM}-\ref{sec:SNM}.  
We also note that the algorithm may converge to a not necessarily suitable solution.
To avoid this problem, $m$ initial points are randomly prepared to produce $m$ solutions obtained from the EEE algorithm. 
Among the $m$ solutions, the best solution is chosen such that it minimizes the trimmed BIC criterion given in Section \ref{sec:BIC}.  
In our implementation, we simply set $m=10$.

The augmented function used in the EE-step can be regarded as a bivariate $S(\bPsi|\bPsi^{\ast})$ with the constraint $\bPsi=\bPsi^{\ast}$, where $S(\bPsi|\bPsi^{\ast})$ is a collection of the augmented estimating equations given by
\begin{equation*}
\begin{split}
&\sum_{i=1}^nu_{ik}(\bPsi^{\ast})\left\{w(\by_i;\bth_k^{\ast})\frac{\partial}{\partial\bth_k}\log f(\by_i;\bth_k)-C(\bth_k)\right\}=0,\\
&\sum_{i=1}^n\frac{u_{ik}(\bPsi^{\ast})}{\pi_k}\frac{w(\by_i;\bth_k^{\ast})}{B(\bth_k^{\ast})}-\sum_{i=1}^n\frac{u_{ik}(\bPsi^{\ast})}{\pi_K}\frac{w(\by_i;\bth_K^{\ast})}{B(\bth_K^{\ast})}=0,
\end{split}
\end{equation*}
for $k=1,\ldots,K$ and $u_{ik}(\bPsi)=E[u_{ik}|\by_i;\bPsi]$.
The EEE algorithm updates the estimate of $\bPsi$ by solving $S(\bPsi|\bPsi^{(s)})=0$ in the $s$th iteration. 
It is assumed that the component distribution $f(\cdot;\bth_k)$ is continuous and differentiable with respect to $\bth_k$.
Since the density weight $w(\cdot;\bth_k)$ inherits these properties, bivariate function $S(\bPsi|\bPsi^{\ast})$ is continuous with respect to $\bPsi$ and $\bPsi^{\ast}$.
Hence, if the sequence of estimators from the EEE algorithm converges, it will converge to the solution of the augmented equation $S(\bPsi|\bPsi)=0$. 
From Lemma 1 in \cite{TQ2016}, the sequence of estimators $\{\bPsi^{(s)}\}_{s=1,2,\ldots}$ that solve $S(\bPsi^{(s+1)}|\bPsi^{(s)})$ converges to the solution $\bPsi^0$ of $S(\bPsi|\bPsi)=0$ in the neighborhood of $(\bPsi^0,\bPsi^0)$ if $\|S_\bPsi^{-1}S_{\bPsi^{\ast}}\|_2<1$,  where $S_\bPsi=\partial S(\bPsi|\bPsi^{\ast})/\partial\bPsi$ and $S_{\bPsi^{\ast}}=\partial S(\bPsi|\bPsi^{\ast})/\partial\bPsi^{\ast}$.

\subsection{Classification and outlier detection}\label{sec:outlier}
The observations are classified into clusters using the posterior probability \eqref{post.prob}. 
We adopt the standard method of classifying an observation into a cluster with the maximum posterior probability. 
It should be noted that the classification is applied to all  observations, both non-outliers and outliers.

Given the cluster assignment, outlier detection is performed based on a fitted robust model. 
Although distance-based outlier detection rules \citep[e.g.][]{CF2011, Greco2020} can be adopted under Gaussian mixture models, such approaches are not necessarily applicable to other general mixture models that we are concerned with. 
Here, we propose an alternative outlier detection method that uses the component density (or probability mass) function $f(\by_i;\bth_k)$.
Suppose the $i$th observation is classified into the $k$th cluster. 
We define $q_i=f(\by_i;\widehat{\bth}_k)$ as the value of the density function of $\by_i$ under the $k$th cluster.  
Our strategy computes the probability ${\rm pr}(q_i;\bth_k)={\rm P}(f(Y;\bth_k)\leq q_i)$ where $Y$ is a random variable with a density $f(\cdot;\bth_k)$.
Under a general $f$, this probability is not analytically available, but we can approximately compute ${\rm pr}(q_i;\widehat{\bth}_k)$ using the Monte Carlo approximation given by 
$$
{\rm pr}(q_i;\widehat{\bth}_k) \approx R^{-1}\sum_{r=1}^R I\{f(Y^{(r)};\bth_k)\leq q_i\},
$$ 
where $R$ is the number of Monte Carlo samples and $Y^{(r)}$ is the $r$th Monte Carlo sample generated from the distribution $f(\cdot;\bth_k)$.  
Note that a smaller value of ${\rm pr}(q_i;\widehat{\bth}_k)$ means that the $i$th observation is more likely to be an outlier. 
Then, our outlier detection strategy declares the $i$th observation as an outlier when ${\rm pr}(q_i;\widehat{\bth}_k)\leq \alpha$ for some fixed $\alpha$.
Note that this approach may result in both type I error (a non-outlier wrongly flagged as an outlier) and type II error (a genuine outlier is not flagged), and larger $\alpha$ leads to the larger likelihood of the type I error. 
We here consider $\alpha\in\{0.001, 0.005, 0.01\}$ following the popular choices of the thresholding probability values in the distance-based outlier detection.

\subsection{Selection of the tuning parameters}\label{sec:BIC}
In practice, the number of components $K$ is unknown and must be chosen reasonably. 
When the data contains outliers, they should not affect selecting $K$; otherwise the selected $K$ can be different from the true $K$. 
Here, we utilize the results of the outlier detection discussed in the previous section.
Let $S_{\alpha}$ be the index set of observations flagged as non-outliers under thresholding level $\alpha$. 
We then propose the trimmed BIC criterion as follows: 
\begin{equation}\label{SBIC}
{\rm BIC}(K)=-\frac{2n}{|S_{\alpha}|}\sum_{i\in S_{\alpha}}\log f_M(\by_i;\widehat{\bPsi}) + |\bPsi|\log n,
\end{equation}
where $|S_{\alpha}|$ and $|\bPsi|$ are the cardinalities of $S_{\alpha}$ and $\bPsi$, respectively.  
If $|S_{\alpha}|=n$ (no outliers are detected), then the criterion is reduced to the standard BIC criterion. 
The optimal $K$ value is selected as the minimizer of this criterion.

Regarding the choice of $\gamma$, we first note that the proposed EEE algorithm reduces to the standard EM algorithm when $\gamma=0$. 
With a larger $\gamma$, the observations with the small component densities are more strongly downweighted by the density weights $w(\by_i;\bth_k)$. 
Thus, the proposed EEE algorithm becomes more robust against outliers. 
However, when there are no outliers, using a large value of $\gamma$ may result in a loss of efficiency by unnecessarily downweighting non-outliers. 
Specifically, the use of $\gamma$ greater than $0.5$ may lead to considerable efficiency loss and numerical instability, where this phenomenon is partly demonstrated under the simple models in \cite{Basu1998}. 
Hence, we suggest using a small positive value for $\gamma$. 
Specifically, we recommend using $\gamma=0.2$ or $0.3$ as in our numerical studies in Sections \ref{sec:sim} and \ref{sec:app}. 
Another practical idea is to monitor the trimmed BIC (\ref{SBIC}) for several values of $\gamma$ and $K$ as demonstrated in Section \ref{sec:swiss}.

\subsection{Asymptotic variance-covariance matrix}
Here, the asymptotic variance-covariance matrix of the estimator is considered. 
Let $G_i^{\ast}(\bPsi|z_i)$ denote the complete estimating functions for the $i$th observation given in \eqref{WEE} and let $G_i(\bPsi)=E[G_i^{\ast}(\bPsi|z_i)]$ denote the estimating equations in which the unobserved $z_i$ (or $u_{ik}$) is imputed with its conditional expectation. 
Furthermore, let $\widehat{\bPsi}$ denote the estimator which is the solution of $\sum_{i=1}^nG_i(\widehat{\bPsi})=0$. 
Note that the augmented estimating equations are unbiased since the complete estimating equations \eqref{WEE} are unbiased.
Then, under some regularity conditions, the asymptotic distribution of $\widehat{\bPsi}$ is $\sqrt{n}(\widehat{\bPsi}-\bPsi)\to N(0, V_{\gamma})$ where the asymptotic variance-covariance matrix $V_\gamma$ is given by 
\begin{align*}
V_{\gamma}&=\lim_{n\to\infty} \left(\frac1n\sum_{i=1}^n\frac{\partial G_i(\bPsi)}{\partial\bPsi}\right)^{-1}\left\{\frac1n\sum_{i=1}^n{\rm Var}(G_i(\bPsi))\right\}\left(\frac1n\sum_{i=1}^n\frac{\partial G_i(\bPsi)^{\top}}{\partial\bPsi}\right)^{-1}.
\end{align*}
This can be consistently estimated by replacing ${\rm Var}(G_i(\bPsi))$ with $G_i(\widehat{\bPsi})G_i(\widehat{\bPsi})^{\top}$ and $\partial G_i(\bPsi)/\partial\bPsi$ with $\partial G_i(\bPsi)/\partial\bPsi|_{\bPsi=\widehat{\bPsi}}$. 
In practice, it is difficult to obtain an analytical expression for the derivative of $G_i(\bPsi)$.  
Therefore, the derivative is numerically computed by using the outputs of the EEE algorithm.  
Let $\bPsi^{(s)}$ denote the final estimate of the EEE algorithm. 
Then the derivative $\partial G_i(\bPsi)/\partial\bPsi_j$ for $j=1,\ldots,{\rm dim}(\bPsi)$ evaluated at $\widehat{\bPsi}$ can be numerically approximated by $\{G_i(\bPsi^{(s)})-G_i(\bPsi^{\ast}_{(j)})\}/(\bPsi_j^{(s)}-\bPsi_j^{(s-1)})$, where $\bPsi^{\ast}_{(j)}$ is obtained by replacing the $\bPsi_j^{(s)}$ in $\bPsi^{(s)}$ with $\bPsi_j^{(s-1)}$.

\section{Robust Gaussian mixture}\label{sec:GM}

The Gaussian mixture models are the most famous and widely adopted mixture models for model-based clustering, even though the performance can be severely affected by outliers due to the light tails of the Gaussian (normal) distribution.
Here, a new approach to the robust fitting of the Gaussian mixture model using the proposed weighted complete estimating equations is presented. 
Let $f(\by_i;\bth_k)=\phi_p(\by_i;\bmu_k,\bSig_k)$ denote the $p$-dimensional normal density $\phi_p$. 
It holds that
$$
B(\bth_k)=|\bSig_k|^{-\gam/2}(2\pi)^{-p\gam/2}(1+\gamma)^{-p/2},
$$
and $C(\bth_k)=(\textbf{0}_p, {\rm vec}(C_{k2}))$, where $\textbf{0}_p$ is a $p$-dimensional vector of $0$ and  
\begin{align*}
C_{k2}=-\frac{\gamma}{2} |\bSig_k|^{-\gam/2}\bSig_k^{-1}(2\pi)^{-p\gam/2}(1+\gamma)^{-p/2-1}.
\end{align*}
Then, the weighted complete estimating equations for $\bmu_k$ and $\bSig_k$ are given by
\begin{align*}
&\sum_{i=1}^nu_{ik}w(\by_i;\bth_k)(\by_i-\bmu_k)=0\\
&\sum_{i=1}^nu_{ik}\Big\{w(\by_i;\bth_k)\bSig_k-w(\by_i;\bth_k)(\by_i-\bmu_k)(\by_i-\bmu_k)^{\top}\Big\}-g(\bSig_k)\Big(\sum_{i=1}^nu_{ik}\Big)\bSig_k=0,
\end{align*}
for $k=1,\ldots,K$, where $g(\bSig_k)=\gamma|\bSig_k|^{-\gamma/2}(2\pi)^{-p\gam/2}(1+\gamma)^{-p/2-1}$.
Hence, starting from some initial values $\bPsi^{(0)}=(\bpi^{(0)},\bth_1^{(0)},\ldots,\bth_K^{(0)})$, the proposed  EEE algorithm repeats the following two steps in the $s$th iteration:

\begin{itemize}
\item[-]
{\it E-step:} \ \ 
The standard E-step is left unchanged as 
$$
u_{ik}^{(s)}=\frac{\pi_k^{(s)}\phi_p(\by_i;\bmu_k^{(s)},\bSig_k^{(s)})}{\sum_{\ell=1}^K\pi_\ell^{(s)}\phi_p(\by_i;\bmu_\ell^{(s)},\bSig_\ell^{(s)})}.
$$

\item[-]
{\it EE-step:} \ \ 
Compute the density weight $w_{ik}^{(s)}=w(\by_i;\bth_k^{(s)})$ and then update the membership probabilities $\pi_k$'s as described in Section \ref{sec:WEE}. 
The component-specific parameters $\theta_k$'s are updated as follows:
\begin{align*}
&\bmu_k^{(s+1)}=\frac{\sum_{i=1}^n u_{ik}^{(s)}w_{ik}^{(s)} \by_i}{\sum_{i=1}^n u_{ik}^{(s)}w_{ik}^{(s)}},\\ 
&\bSig_k^{(s+1)}=\frac{\sum_{i=1}^n u_{ik}^{(s)} w_{ik}^{(s)}(\by_i-\bmu_k^{(s+1)})(\by_i-\bmu_k^{(s+1)})^{\top} }{\sum_{i=1}^n u_{ik}^{(s)}w_{ik}^{(s)}-g(\bSig_k^{(s)})\sum_{i=1}^n u_{ik}^{(s)}},
\end{align*}
and the mixing proportion $\pi_k$ is updated as \eqref{pi-update}. 
\end{itemize}

Note that all the formulas in the above steps are obtained in closed form.
This is one of the attractive features of the proposed method, as it does not require any computationally intensive methods such as Monte Carlo integration.

To choose reasonable starting values $\bPsi^{(0)}$, we employ existing robust estimation methods of Gaussian mixture models or clustering, such as trimmed robust clustering \citep[TCL;][]{tclust} and improper maximum likelihood \citep[IML;][]{PC2016}.

To stabilize the estimation of the variance-covariance matrix $\bSig_k$, it would be beneficial to impose the following eigen-ratio constraint:
\begin{equation}\label{eigen}
\frac{\max_{j=1,\ldots,p}\max_{k=1,\ldots,K} \lambda_j(\bSig_k)}{\min_{j=1,\ldots,p}\min_{k=1,\ldots,K} \lambda_j(\bSig_k)}\leq c,
\end{equation}
where $\lambda_j(\bSig_k)$ denotes the $j$th eigenvalue of the covariance matrix $\bSig_k$ in the $k$th component and $c$ is a fixed constant. 
When $c=1$, a spherical structure is imposed on $\bSig_k$, and a more flexible structure is allowed under a large value of $c$.
To reflect the eigen-ratio constraint in the EEE algorithm, the eigenvalues of $\bSig_k^{(s)}$ can be simply replaced with the truncated version $\lambda_j^{\ast}(\bSig_k)$ where $\lambda_j^{\ast}(\bSig_k)=c$ if $\lambda_j(\bSig_k)>c$, $\lambda_j^{\ast}(\bSig_k)=\lambda_j(\bSig_k)$ if $c\theta_c\leq \lambda_j(\bSig_k)\leq c$ and $\lambda_j^{\ast}(\bSig_k)=c\theta_c$ if $\lambda_j(\bSig_k)<c\bth_c$, where $\theta_c$ is an unknown constant that depends on $c$. 
The procedure of \cite{Fritz2013} is employed in this study.   
Note that if $c$ is set to a small value, the constraint for $\bSig_k$ is too severe such that the solution of the weighted estimating equation might not exist.

\section{Robust mixture of experts}\label{sec:MOE}
A mixture of experts model \citep{Jacob1991,Mc2004} is a useful tool for modelling nonlinear regression relationships.
Models that include a simple mixture of normal regression models and their robust versions have been proposed in the literature \citep[e.g.][]{Bai2012,Song2014}.
In addition, the robust versions of a mixture of experts based on the non-normal distributions were considered by \cite{Nguyen2016}, and \cite{CH2016}. 
The general model is described as
$$
f_M(y_i|\bx_i;\bPsi)=\sum_{k=1}^K g(\bx_i;\bEta_k)f(y_i|\bx_i;\bth_k),
$$
where $\bx_i$ is a vector of covariates including an intercept term and $g(\bx_i;\bEta_k)$ is the mixing proportion such that $\sum_{k=1}^Kg(\bx_i;\bEta_k)=1$.
For the identifiability of the parameters,  $\bEta_K$ is assumed to be an empty set, since $g(\bx_i;\bEta_K)$ is completely determined by $\bEta_1,\ldots,\bEta_{K-1}$. 
A typical form of the continuous response variables adopts $f(y_i|\bx_i;\bth_k)=\phi(y_i; \bx_i^{\top}\bbe_k, \sigma_k^2)$ and $g(\bx_i;\bEta_k)=\exp(\bx_i^{\top}\bEta_k)/\sum_{k=1}^K\exp(\bx_i^{\top}\bEta_k)$. 
Let $\bPsi$ be the set of the unknown parameters, $\bEta_k$, and $\bth_k$. 
Compared to the standard mixture model (\ref{Mix}), the mixing proportion $g(\bx_i;\bEta_k)$ is a function of $\bx_i$ parameterized by $\bEta_k$.

As before, the latent variable $z_i$ such that $P(z_i=k)=g(\bx_i;\bEta_k)$ for $k=1,\ldots,K$ is introduced. 
Then, the complete weighted estimating equations are given by
\begin{equation*}\label{MOE-WEE}
\begin{split}
&\sum_{i=1}^nu_{ik}\left\{w(y_i;\bth_k)\frac{\partial}{\partial\bth_k}\log f(y_i;\bth_k)-C_i(\bth_k)\right\}=0,\\
&\sum_{i=1}^n\frac{u_{ik}}{g(\bx_i;\bEta_k)}\frac{\partial g(\bx_i;\bEta_k)}{\partial\bEta_k}\frac{w(y_i;\bth_k)}{B_i(\bth_k)}-\sum_{i=1}^n \frac{u_{iK}}{g(\bx_i;\bEta_K)}\frac{\partial g(\bx_i;\bEta_k)}{\partial\bEta_k}\frac{w(y_i;\bth_K)}{B_i(\bth_K)}=0,
\end{split}
\end{equation*}
where 
\begin{align*}
&C_i(\bth_k)
=\int f(t|\bx_i;\bth_k)w(t;\bth_k)\frac{\partial}{\partial\bth_k}\log f(t|\bx_i;\bth_k)dt,\\
& \ \ \ \ 
B_i(\bth_k)
=\int f(t|\bx_i;\bth_k)w(t;\bth_k)dt.
\end{align*}
Starting from some initial values $\bPsi^{(0)}$, the proposed EEE algorithm repeats the following two steps in the $s$th iteration:
\begin{itemize}
\item[-]
{\it E-step:} \ \ 
The standard E-step is left unchanged as
$$
u_{ik}^{(s)}=\frac{g(\bx_i;\bEta_k^{(s)})f(y_i|\bx_i;\bth_k^{(s)})}{\sum_{\ell=1}^Kg(\bx_i;\bEta_\ell^{(s)})f(y_i|\bx_i;\bth_\ell^{(s)})}.
$$

\item[-]
{\it EE-step:} \ \ 
Update $\bEta_k$ and $\bth_k$ by solving the following equations:
\begin{equation}\label{MOE-EE}
\begin{split}
&\sum_{i=1}^nu_{ik}^{(s)}\left\{w(y_i;\bth_k^{(s)})\frac{\partial}{\partial\bth_k}\log f(y_i;\bth_k)-C_i(\bth_k)\right\}=0,\\
&\sum_{i=1}^n\frac{u_{ik}^{(s)}}{g(\bx_i;\bEta_k)}\frac{\partial g}{\partial\bEta_k}\frac{w(y_i;\bth_k^{(s+1)})}{B_i(\bth_k^{(s+1)})}-\sum_{i=1}^n\frac{u_{iK}^{(s)}}{g(\bx_i;\bEta_K)}\frac{\partial g}{\partial\bEta_k}\frac{w(y_i;\bth_K^{(s+1)})}{B_i(\bth_K^{(s+1)})}=0.
\end{split}
\end{equation}
\end{itemize}

When the mixture components are the normal linear regression models given by $f(y_i|\bx_i;\bth_k)=\phi(y_i; \bx_i^{\top}\bbe_k, \sigma_k^2)$ with $\bth_k=(\bbe_k^{\top},\sigma_k^2)$, the bias correction terms $C_i(\bth_k)\equiv(\textbf{0}_p,C_{i2}(\bth_k))$ and $B_i(\bth_k)$ can be analytically obtained as 
$$
C_{i2}(\bth_k)=-(\gamma/2)(\sigma_k^2)^{-1-\gamma/2}(2\pi)^{-\gamma/2}(1+\gamma)^{-3/2}
$$
$$
B_i(\bth_k)=(2\pi\sigma_k^2)^{-\gamma/2}(1+\gamma)^{-1/2}.
$$
The first equation in \eqref{MOE-EE} can be solved analytically, and the closed-form updating steps similar to those in Section \ref{sec:GM} can be obtained. 
On the other hand, the second equation is a function of $\bEta_1,\ldots,\bEta_{K-1}$ and cannot be solved analytically.
Note that the solution corresponds to the maximizer of the weighted log-likelihood function of the multinomial distribution given by
$$
\sum_{i=1}^n\sum_{k=1}^K u_{ik}^{(s)}\frac{w(\by_i;\bth_k^{(s+1)})}{B_i(\bth_k^{(s+1)})}\log g(\bx_i;\bEta_k),
$$ 
since its first-order partial derivatives with respect to $\bEta_k$ are reduced to the second equation in (\ref{MOE-EE}).
Thus, the updating step for $\bEta_k$ can be readily carried out.

In setting the initial values of the EEE algorithm, we first randomly split the data into $K$ groups and apply some existing robust methods to estimate $\bth_k$ for $k=1,\ldots,K$, which can be adopted for the initial values $\bth_k^{(0)}$ of $\bth_k$.
For example, when $f(\cdot;\bth_k)$ is the normal linear regression as adopted in Section \ref{sec:tone}, we employ an M-estimator available in the \verb+rlm+ function in R package `MASS' \citep{VR2002}. 
For the initial values of $\bEta_k$, we suggest simply using $\bEta_k^{(0)}=(0,\ldots,0)$.

\section{Robust skew normal mixture}\label{sec:SNM}

We next consider the use of the $p$-dimensional skew normal distribution \citep{Azzalini96} for the $k$th component. 
The mixture model based on skew normal distributions is more flexible than a multivariate Gaussian mixture, especially when the cluster-specific distributions are not symmetric but skewed.
Several works have been conducted on the maximum likelihood estimation of a skew normal mixture \citep{Lin2007,Lin2009}. 
Despite the flexibility in terms of skewness, because the skew normal distribution still has light tails as the normal distribution, the skew normal mixture is also sensitive to outliers. 
Alternative mixture models using heavy-tailed and skewed distributions have also been proposed in the literature \citep[e.g.][]{Lee2017,Morris2019}. 
Here, we consider the robust fitting of the skew normal mixture within the proposed framework.

We first note that the direct application of the skew normal distribution to the proposed EEE algorithm would be computationally intensive since the bias correction term $C(\bth_k)$ cannot be obtained analytically, and a Monte Carlo approximation would be required in each iteration.
Instead, we provide a novel algorithm that exploits the stochastic representation of the multivariate skew normal distribution.
The resulting algorithm is a slight extension of the proposed algorithm in Section \ref{sec:WEE}. 
\cite{Schnatter2010} provided the following hierarchical representation for $\by_i$ given $z_i$:  
\begin{equation}\label{Hie}
\begin{split}
&\by_i|(z_{i}=k) = \bmu_k + \bpsi_k v_{ik} + \ep_{ik}, \\
&\epsilon_{ik}\sim N(0,\bSig_k), \quad v_{ik}\sim N^{+}(0,1),
\end{split}
\end{equation}
where $\bmu_k$ is the $p\times1$ vector of location parameters, $\bpsi_k$ is the $p\times1$ vector of the skewness parameters, and $\bSig_k$ is the covariance matrix. 
Here, $N^{+}(a,b)$ denotes the truncated normal distribution on the positive real line with mean and variance parameters $a$ and $b$, respectively, and its the density function $\phi^+(x;a,b^2)$ is given by 
\begin{align*}
&\phi^+(x;a,b)=\frac{1}{\Phi(a/b)}\phi(x;a,b)I(x\geq 0),
\end{align*}
where $\Phi(\cdot)$ is the distribution function of the standard normal distribution. 
By defining 
$$
\bOmega_k = \bSig_k + \bpsi_k\bpsi_k^{\top},\quad \bal_k = \frac{\bomega_k\bOmega_k^{-1}\bpsi_k}{\sqrt{1-\bpsi_k^{\top}\bOmega_k^{-1}\bpsi_k}},
$$
where $\bomega_k=(\bOmega_{11}^{1/2},\dots,\bOmega_{pp}^{1/2})$. 
The probability density function of the multivariate skew normal distribution of \cite{Azzalini96} is given by
\begin{equation}\label{SND}
f_{SN}(\by;\bmu_k,\bOmega_k,\bal_k)=2\phi_p(\by;\bmu_k,\bOmega_k)\Phi(\bal_k^{\top}\bomega^{-1}_k(\by-\bmu_k)). 
\end{equation}

From \eqref{Hie}, the conditional distribution of $\by_i$ given both $z_i$ and $v_{ik}$ is normal, namely, $\by_i|(z_i=k),v_{ik}\sim N(\bmu_k + \bpsi_k v_{ik},\bSig_k)$, so that the bias correction terms under given $z_i$ and $v_{ik}$ can be easily obtained in the same way as in the case of the Gaussian mixture models. 
Therefore, we consider the following complete estimating equations for the parameters conditional on both $z_i$ and $v_{ik}$:
\begin{equation*}
\begin{split}
&\sum_{i=1}^nu_{ik}w_{ik}(\by_i-\bmu_k-\bpsi_k v_{ik})=0,\\
&\sum_{i=1}^nu_{ik}w_{ik}\Big\{\bSig_k-(\by_i-\bmu_k-\bpsi_kv_{ik})(\by_i-\bmu_k-\bpsi_kv_{ik})^{\top}\Big\}-g(\bSig_k)\left(\sum_{i=1}^nu_{ik}\right)\bSig_k=0,\\
&\sum_{i=1}^nu_{ik}v_{ik}w_{ik}(\by_i-\bmu_k-\bpsi_k v_{ik})=0,\\
\end{split}
\end{equation*}
where $w_{ik}\equiv w(\by_i;\bth_k)$ and $g(\bSig_k)$ have the same forms as in the Gaussian mixture case.

The proposed EEE algorithm for the robust fitting of the skew normal mixture is obtained after some modification to the normal case. 
Since the complete estimating equations contain the additional latent variables $v_{ik}$, some additional steps are included to impute the moments of $v_{ik}$ in the equations. 
The moments are those of the conditional posterior distribution of $v_{ik}$, given $z_i=k$, which is $N^+(\delta_{ik},\tau_{ik}^2)$, where 
\begin{equation*}
\delta_{ik}=\frac{\bpsi_k^{\top} \bSig_k^{-1}(\by_i-\bmu_k)}{\bpsi_k^{\top}\bSig_k^{-1}\bpsi_k+1}, \ \ \ \ 
\tau_{ik}^2=\frac{1}{\bpsi_k^{\top}\bSig_k^{-1}\bpsi_k+1}.
\end{equation*}
We define the following quantities:
\begin{equation}\label{SN-E}
\begin{split}
&t^2_{ik}=\left(\gam\bpsi_k^{\top}\bSig_k^{-1}\bpsi_k + \frac{1}{\tau^2_{ik}}\right)^{-1}, \ \ \ \ 
m_{ik}=t^2_{ik}\left\{\gam\bpsi_k^{\top}\bSig_k^{-1}(\by_i-\bmu_k) + \frac{\delta_{ik}}{\tau_{ik}^2}\right\}, \\
&U_{ik}=|\bSig_k|^{-\gam/2}(2\pi)^{-\gam p/2}\frac{\Phi(m_{ik}/t_{ik})}{\Phi(\delta_{ik}/\tau_{ik})}\left(\frac{t_{ik}^2}{\tau_{ik}^2}\right)^{1/2} \\\
& \times\exp\left\{-\frac{\gamma}2(\by_i-\bmu_k)^{\top}\bSig_k^{-1}(\by_i-\bmu_k)-\frac{\delta^2_{ik}}{2\tau^2_{ik}}+\frac{m_{ik}^2}{2t_{ik}^2}\right\}.
\end{split}
\end{equation}
Note that we have for $j=0,1$ and $2$
\begin{align*}
E[u_{ik}w_{ik}v_{ik}^j|\by_i]
&=E[E[u_{ik}w_{ik}v_{ik}^j|\by_i,u_{ik}]|\by_i]=E[u_{ik}|\by_i]E[w_{ik}v_{ik}^j|u_{ik},\by_i].
\end{align*}
Since the conditional distribution of $v_{ik}$ given $(u_{ik},\by_i)$ is $N^+(\delta_{ik},\tau_{ik}^2)$, we have 
\begin{align*}
&E[v_{ik}^jw_{ik}|u_{ik},\by_i]\\
&=\int_0^\infty |\bSig_k|^{-\gam/2}(2\pi)^{-\gam p/2}\exp\left\{-\frac{\gamma}{2}(\by_i-\bmu_k-v_{ik}\bpsi_k)^{\top}\bSig_k^{-1}(\by_i-\bmu_k-v_{ik}\bpsi_k) \right\}\\
& \ \ \ \ \ \times \frac{v_{ik}^j}{\Phi(\delta_{ik}/\tau_{ik})}(2\pi \tau_{ik}^2)^{-1/2}\exp\left\{-\frac{(v_{ik}-\delta_{ik})^2}{2\tau_{ik}^2}\right\}d v_{ik}\\
&=|\bSig_k|^{-\gam/2}(2\pi)^{-\gam p/2}\frac{\Phi(m_{ik}/t_{ik})}{\Phi(\delta_{ik}/\tau_{ik})}\left(\frac{t_{ik}^2}{\tau_{ik}^2}\right)^{1/2}\int_0^{\infty}v_{ik}^j\phi_+(v_{ik};m_{ik},t_{ik}^2)dv_{ik}\\
& \ \ \ \ \ \ \times\exp\left\{-\frac{\gamma}2(\by_i-\bmu_k)^{\top}\bSig_k^{-1}(\by_i-\bmu_k)-\frac{\delta^2_{ik}}{2\tau^2_{ik}}+\frac{m_{ik}^2}{2t_{ik}^2}\right\},
\end{align*}
and it follows from \cite{Lin2007} that 
\begin{align*}
&\int_0^{\infty}v_{ik}^1\phi_+(v_{ik};m_{ik},t_{ik}^2)dv_{ik}
=m_{ik}+t_{ik}\frac{\phi(m_{ik}/t_{ik})}{\Phi(m_{ik}/t_{ik})},\\
&\int_0^{\infty}v_{ik}^2\phi_+(v_{ik};m_{ik},t_{ik}^2)dv_{ik}=m_{ik}^{2}+t_{ik}^{2} + m_{ik}t_{ik}\frac{\phi(m_{ik}/t_{ik})}{\Phi(m_{ik}/t_{ik})}. 
\end{align*}
These expressions are applied to the analytical updating steps in E-step.
Starting from some initial values of the parameters, $\bPsi^{(0)}=(\bpi^{(0)},\bth_1^{(0)},\ldots,\bth_K^{(0)})$, the proposed EEE algorithm iteratively updates the parameters in the $s$th iteration as follows:

\begin{itemize}

\item[-]
{\it E-step:} \ \ 
Compute the posterior expectations: 
\[
\begin{split}
u_{ik}^{(s)}
&\equiv E^{(s)}[u_{ik}|\by_i]=\frac{\pi_k^{(s)}f_{SN}(\by_i;\bmu_k^{(s)},\bOmega_k^{(s)},\bal_\ell^{(s)})}{\sum_{\ell=1}^K\pi_\ell^{(s)}f_{SN}(\by_i;\bmu_\ell^{(s)},\bOmega_\ell^{(s)},\bal_\ell^{(s)})},\\
V_{ik}^{0(s)}
&\equiv E^{(s)}[w_{ik}|\by_i,u_{ik}]
=U_{ik}^{(s)},\\
V_{ik}^{1(s)}
&\equiv E^{(s)}[w_{ik}v_{ik}|\by_i,u_{ik}]=U_{ik}^{(s)}\left\{m_{ik}^{(s)}+t_{ik}^{(s)}\frac{\phi(m_{ik}^{(s)}/t_{ik}^{(s)})}{\Phi(m_{ik}^{(s)}/t_{ik}^{(s)})}\right\},\\
V_{ik}^{2(s)}
&\equiv E^{(s)}[w_{ik}v^2_{ik}|\by_i,u_{ik}]=U_{ik}^{(s)}\left\{m_{ik}^{2(s)}+t_{ik}^{2(s)} + m_{ik}^{(s)}t_{ik}^{(s)}\frac{\phi(m_{ik}^{(s)}/t_{ik}^{(s)})}{\Phi(m_{ik}^{(s)}/t_{ik}^{(s)})}\right\}, 
\end{split}
\]
where $m_{ik}^{(s)}$ and $t_{ik}^{(s)}$ stand for $m_{ik}$ and $t_{ik}^2$ given in \eqref{SN-E}, respectively, evaluated with the current parameter values.

\item[-]
{\it EE-step:} \ \ 
Update the membership probabilities $\pi_k$'s as in Section \ref{sec:WEE} and component-specific parameters $\bth_k$'s as
\begin{align*}
&\pi_k^{(s+1)}=\frac{\sum_{i=1}^nu_{ik}^{(s)}V_{ik}^{0(s)}/B(\bth_k^{(s)})}{\sum_{\ell=1}^K\sum_{i=1}^nu_{i\ell}^{(s)}V_{i\ell}^{0(s)}/B(\bth_\ell^{(s)})},\\
&\bmu_k^{(s+1)}=\frac{\sum_{i=1}^n u_{ik}^{(s)}(V_{ik}^{0(s)}\by_i-V_{ik}^{1(s)}\bpsi_k^{(s)})}{\sum_{i=1}^n u_{ik}^{(s)}V_{ik}^{0(s)}},\\ 
&\bpsi_k^{(s+1)}=
\frac{\sum_{i=1}^nu_{ik}^{(s)}V_{ik}^{1(s)}(\by_i-\bmu_k^{(s+1)})}
{\sum_{i=1}^n u_{ik}^{(s)}V_{ik}^{2(s)}}, \\
&\bSig_k^{(s+1)}
=\frac{\sum_{i=1}^n u_{ik}^{(s)} \bEta_{ik}^{(s,s+1)}}
{\sum_{i=1}^n u_{ik}^{(s)}V_{ik}^{0(s)}-g\big(\bSig_k^{(s)}\big)\sum_{i=1}^n u_{ik}^{(s)}},
\end{align*}
for $k=1,\ldots,K$ where 
\begin{equation*}
\begin{split}
\bEta_{ik}^{(s,s+1)}=&V_{ik}^{0(s)}(\by_i-\bmu_k^{(s+1)})(\by_i-\bmu_k^{(s+1)})^{\top}-V_{ik}^{1(s)}\bpsi_k^{(s+1)}(\by_i-\bmu_k^{(s+1)})^{\top}\\
&-V_{ik}^{1(s)}(\by_i-\bmu_k^{(s+1)})(\bpsi_k^{(s+1)})^{\top}+V_{ik}^{2(s)}\bpsi^{(s+1)}_k(\bpsi_k^{(s+1)})^{\top}.
\end{split}
\end{equation*}
\end{itemize}

Note that the form of $B(\bth_k)$ under the skew normal mixture is the same as that in the case of the normal mixture, because it holds that $E[w(\by_i;\bth_k)]=E[E[w(\by_i;\bth_k)|u_{ik}]]$ and the conditional distribution of $\by_i$ given $u_{ik}$ is the multivariate normal with the variance-covariance matrix $\bSig_k$.
In addition, note that all the steps in the above algorithm are obtained in closed forms by successfully exploiting the stochastic representation of the skew normal distribution in the complete weighted complete estimating equations.
In our EEE algorithm, we impose the eigen-ratio constraint considered in Section \ref{sec:GM} to avoid an ill-posed problem.

The following strategy is employed to set the initial values of the algorithm.
First, an existing robust algorithm for fitting Gaussian mixture models is applied to obtain the cluster assignment and mean vectors of the clusters. 
We then set $\bmu_k^{(0)}$ as the estimated mean vector and set $\bpsi_k^{(0)}$ to be the vector of marginal sample skewness of the observations in the $k$th cluster. 
For $\bSig_k$ and $\pi_k$, we set $\bSig_k^{(0)}=I_p$ and $\pi_k=1/K$.

\section{Simulation study}\label{sec:sim}

\subsection{Gaussian mixture}\label{sec:sim-GM}
The performance of the proposed method for the robust fitting of a Gaussian mixture is evaluated along with existing methods through simulation studies.
First, the performance in terms of the estimation accuracy of the unknown model parameters is examined.
We set the underlying true distribution to be a $p$-dimensional Gaussian mixture model with $K=3$ components, where the parameters in each Gaussian distribution are given by
\begin{align*}
&\bmu_1=(-\xi,-\xi,0,\ldots,0), \ \ 
\bmu_2=(\xi,-\xi,0,\ldots,0),  \ \ \bmu_3=(\xi,\xi,0,\ldots,0)
\end{align*}
for some $\xi>0$ and $\bSig_k={\rm blockdiag}(\bSig_k^{\ast},I_{p-2})$ with 
\begin{align*}
&\bSig_1=\left(\begin{array}{cc}
2 & 0.3 \\
0.3 & 1 
\end{array}\right), \ \ 
\bSig_2=\left(\begin{array}{cc}
1 & -0.3 \\
-0.3 & 1 
\end{array}\right), \ \ 
\bSig_3=\left(\begin{array}{cc}
1 & 0.3 \\
0.3 & 2 
\end{array}\right).
\end{align*}
$\xi$ controls the separation of the three clusters.
The following two scenarios, $\xi=2$ and $\xi=3$, correspond to overlapping and well-separated clusters, respectively.
In this study, the sample size is set to $n=500$, and the three cases of the dimension, $p\in \{2, 5, 10\}$, are considered (the results under $n=2000$ are provided in the Supplementary Material).
The true mixing probabilities are fixed at $\pi_1=0.3$, $\pi_2=0.3$ and $\pi_3=0.4$. 
Outliers are generated from the uniform distribution on $A\setminus B$ where $A=(-10,10)\times (-5,5)\times (-3,3)^{p-2}$ and $B$ is the collection of points where the minimum Mahalanobis distance to the $k$th cluster mean is smaller than $5p$, namely,
$$
B = \{x \ | \ \min_{k=1,\ldots,K} (x-\bmu_k)^{\top}\bSig_k^{-1}(x-\bmu_k) \leq 5p\}.
$$  
Let $\omega$ denote the proportion of outliers such that $n(1-\omega)$ observations are drawn from the true mixture distribution, whereas the remaining observations are generated as outliers. 
We adopt the following three cases: $\omega\in \{0, 0.03, 0.06\}$. 
Setting $\omega=0$ means that there are no outliers in the data. 
Therefore, the efficiency loss of the robust methods against standard non-robust methods can be assessed.

For each of the 1000 simulated datasets, the proposed weighted complete estimating equation method is applied with two fixed values of $\gamma$, $\gamma=0.2$, and $0.3$. 
Hereafter, they will be denoted as WCE1 and WCE2, respectively. 
For comparison, the standard (non-robust) maximum likelihood method for fitting the Gaussian mixture (GM) using the EM algorithm is also implemented.
For the existing robust contenders, we consider the trimmed robust clustering \citep[TCL;][]{tclust}, improper maximum likelihood \citep[IML;][]{PC2016} and contaminated Gaussian mixture \citep[CGM;][]{Punzo2016}, which are available in R packages `tclust' \citep{Fritz2012}, `otrimle' \citep{CH2019} and `ContaminatedMixt' \citep{Punzo2018}, respectively.
The trimming level in TCL is set to a default value of $0.05$. 
In CGM, the fully structured variance-covariance is used. 
The same eigen-ratio constraint is imposed for all the methods other than CGM by setting $c=10$ in condition \eqref{eigen}, noting that the true eigen-ratios are $2.27$ for the first and third clusters and $1.86$ for the second cluster. 
In WCE1 and WCE2, $\alpha=0.01$ is employed for outlier detection.

The estimation performance of the methods is evaluated based on the squared error given by $\sum_{k=1}^K\|\widehat{\bmu}_k-\bmu_k\|^2$, $\sum_{k=1}^K\|{\rm vec}(\widehat{\bSig}_k)-{\rm vec}(\bSig_k)\|^2$ and $\sum_{k=1}^K(\widehat{\pi}_k-\pi_k)^2$.
We also computed the following integrated squared error to check the overall estimation accuracy: 
\begin{equation}\label{MISE}
\int_{\Re^p} \left\{\sum_{k=1}^K\widehat{\pi}_k f(x;\widehat{\bth}_k)-\sum_{k=1}^K\pi_k f(x;\bth_k)\right\}^2 {\rm d}x,
\end{equation}
where the integral is approximated by Monte Carlo integration by generating 3000 random numbers uniformly distributed on $[-6,6]^2\times [-3,3]^{p-2}$.  
For measuring the classification accuracy, we calculated the classification error defined as $|S|^{-1}\sum_{i\in S}I(\widehat{z}_i\neq z_i)$, where $\widehat{z}_i$ and $z_i$ are the estimated and true cluster assignments, respectively, and $S$ is the set of non-outliers.  
The above quantities were averaged over 1000 simulated detests, which gives the mean squared error (MSE) of the model parameters, the mean integrated squared errors (MISE) of the density function and the mean classification error (MCE) of the cluster assignment.

Tables \ref{tab:sim-GM1}-\ref{tab:sim-GM3} present the MSE, MISE and MCE of the competing models under $p=2, 5$ and $8$, respectively. 
Firstly, the performance of the standard GM is reasonable only in the cases where the clusters are well-separated ($\xi=3$), and there are no outliers ($\omega=0$). 
When the clusters are overlapping ($\xi=2$), and the data does not contain any outliers, the estimation accuracy of GM deteriorates as the dimensionality $p$ increases. 
In fact, in the case of $p=10$, the robust methods, including the proposed method, performed better than GM. 
In the presence of the outliers, $\omega>0$, the estimation accuracy in terms of MSE and MISE of the robust methods appears comparable. 
For the classification accuracy, the order of the performance of the robust methods appears to be case dependent, and they perform comparably overall. 
For example, in the cases where the clusters are well-separated with $p=5$ and $10$, IML and CGM tend to result in smaller MCE than the proposed method. 
On the other hand, when $p=10$ and the clusters are overlapping, the proposed WCE2 resulted in the smallest MCE.

The performance of outlier detection is also evaluated.
We let $\delta_i$ be an indicator of outliers such that $\delta_i=1$ denotes that $\by_i$ is an outlier.  
The false discovery rate (FDR) and power (PW) are given by ${\rm FDR}=\sum_{i\in S}\widehat{\delta}_i/\sum_{i=1}^n \widehat{\delta}_i$ and ${\rm PW}=\sum_{i\in \Omega}\widehat{\delta}_i/n\omega$, where $\Omega$ denotes the set of outliers. 
Table \ref{tab:sim-GM-out} presents the PDR and PW values averaged over the 1000 simulated datasets \ref{tab:sim-GM-out}. 
The table shows that the proposed method performs well in the low-dimensional cases ($p=2$) with achieving a relatively low FDR and high power compared with the contenders.
It is also shown that, as the dimensionality increases, the FDR for the proposed method increases while maintaining high power.

Next, we consider the situation where the number of outliers is close to the minimum size of the clusters. 
Specifically, we set $\omega=0.06$ and the true mixing probability as $\bpi=(0.1, 0.3, 0.6)$.
Using the same data generating process as in the case of $p=2$, MSE, MISE, and MCE of the six methods are evaluated based on the 1000 simulated datasets. 
The result are reported in Table \ref{tab:sim-GM-add}.
In this simulation setting, TCL appears to have produced the best result with the smallest MSE, MISE, and MCE. 
The table shows that some errors under IML and CGM resulted in very large magnitudes.

Finally, we examine the performance in selecting the number of components.
The same simulation scenarios with $p=2$ are considered.
For the simulated datasets, the optimal $K$ from $\{2,\ldots,6\}$ is selected based on the trimmed BIC criterion \eqref{SBIC}.
Since the trimmed BIC can be applied to a variety of robust methods that provide parameter estimates and detect outliers, the proposed methods, as well as the three existing robust methods (TCL, IML, and CGM) are considered in this study. 
The same settings are used for the tuning parameters as in the previous simulation.
In all the scenarios, $\alpha=0.01$ is used for WCE1 and WCE2. 
In addition, the performance is also assessed with $\alpha\in \{0.005, 0.001\}$ to check the sensitivity with respect to $\alpha$ in the most challenging situations where $\omega=0.06$.
Based on the 300 simulated datasets, the selection probabilities for each $K$ are computed. 
The results are presented in Table \ref{tab:sim-K}.
The table shows that when the data contains outliers, the standard GM method tends to select a larger number of components than the truth since the outliers are recognized as a new cluster. 
On the other hand, the proposed methods (WCE1 and WCE2) are highly accurate in selecting the true number of components even in the presence of outliers.
Among the existing robust methods, TCL shows comparable performance with the proposed methods, whereas IML and CGM perform rather poorly.

Recalling that rather advantageous settings are used for the contenders, we can conclude that the overall estimation performance of the proposed method is comparable with the existing methods. 
The results of this study indicate that the proposed method is also a promising approach for the robust estimation of mixture models.

\begin{table}[!htbp]
\caption{Mean squared errors (MSE) of the model parameters, mean integrated squared errors (MISE) of the density estimation, and mean classification error (MCE) of the proposed methods (WCE1 and WCE2), the standard Gaussian mixture (GM), and the existing robust methods (TCL, IML, CGM), based on the 1000 simulated datasets with $p=2$. 
\label{tab:sim-GM1}
}
\begin{center}
{\footnotesize
\begin{tabular}{cccccccccccccccc}
\hline
& MSE & MSE & MSE &&&MSE & MSE & MSE \\
& $\bmu$ & $\bSig$ & $\bpi$ &MISE& MCE & $\bmu$ & $\bSig$ & $\bpi$ &MISE& MCE   \\
& ($\times 10^2$) & &($\times 10^3$)& ($\times 10^6$) & ($\times 10^2$) & ($\times 10^2$) & &($\times 10^3$)& ($\times 10^6$) & ($\times 10^2$) \\
\hline 
&\multicolumn{5}{c}{$(C=2, \ \omega=0)$}  & \multicolumn{5}{c}{$(C=3, \ \omega=0)$}  \\
WCE1 & 7.27 & 0.37 & 1.85 & 3.22 & 4.55 & 5.16 & 0.23 & 1.33 & 2.60 & 1.12 \\
WCE2 & 7.49 & 0.40 & 1.88 & 3.36 & 4.60 & 5.43 & 0.25 & 1.34 & 2.73 & 1.17 \\
GM & 7.09 & 0.34 & 1.82 & 3.09 & 3.91 & 4.92 & 0.22 & 1.33 & 2.49 & 0.24 \\
TCL & 9.04 & 0.82 & 2.62 & 9.35 & 8.69 & 6.01 & 0.60 & 1.65 & 6.55 & 5.03 \\
IML & 7.64 & 0.61 & 2.37 & 9.44 & 7.18 & 5.23 & 0.41 & 1.60 & 6.60 & 2.47 \\
CGM & 7.15 & 0.47 & 1.84 & 5.07 & 3.91 & 4.94 & 0.33 & 1.33 & 4.06 & 0.24 \\
\hline 
&\multicolumn{5}{c}{$8C=2, \ \omega=0.03)$}  & \multicolumn{5}{c}{$(C=3, \ \omega=0.03)$}   \\
WCE1 & 8.90 & 0.91 & 2.15 & 3.77 & 4.54 & 5.66 & 0.42 & 1.34 & 2.98 & 0.90 \\
WCE2 & 8.28 & 0.57 & 2.00 & 3.59 & 4.56 & 5.77 & 0.34 & 1.35 & 2.91 & 1.02 \\
GM & 47.63 & 17.28 & 10.01 & 13.68 & 6.28 & 13.13 & 6.00 & 1.39 & 11.25 & 0.28 \\
TCL & 8.47 & 0.56 & 2.24 & 6.09 & 5.93 & 5.76 & 0.33 & 1.49 & 3.91 & 2.16 \\
IML & 8.73 & 0.64 & 2.22 & 6.62 & 6.04 & 5.74 & 0.45 & 1.62 & 5.45 & 2.14 \\
CGM & 16.17 & 2.77 & 3.71 & 5.13 & 4.41 & 5.69 & 0.48 & 1.40 & 3.60 & 0.26 \\
\hline 
&\multicolumn{5}{c}{$(C=2, \ \omega=0.06)$} & \multicolumn{5}{c}{$(C=3, \ \omega=0.06)$}  \\
WCE1 & 17.13 & 6.72 & 4.23 & 6.37 & 4.99 & 6.40 & 1.54 & 1.46 & 4.45 & 0.74 \\
WCE2 & 9.77 & 1.36 & 2.33 & 4.12 & 4.63 & 6.02 & 0.60 & 1.45 & 3.36 & 0.90 \\
GM & 164.50 & 52.31 & 32.37 & 24.63 & 10.97 & 32.90 & 23.56 & 1.81 & 20.84 & 0.32 \\
TCL & 9.73 & 0.69 & 2.15 & 4.30 & 4.13 & 6.24 & 0.47 & 1.51 & 3.22 & 0.35 \\
IML & 9.35 & 0.80 & 2.52 & 7.91 & 6.86 & 6.12 & 0.58 & 1.82 & 7.33 & 3.20 \\
CGM & 58.41 & 15.18 & 14.10 & 8.77 & 6.23 & 6.41 & 1.08 & 1.67 & 4.99 & 0.26 \\
\hline
\end{tabular}
}
\end{center}
\end{table}

\begin{table}[!htbp]
\caption{Mean squared errors (MSE) of the model parameters, mean integrated squared errors (MISE) of the density estimation, and mean classification error (MCE) of the proposed methods (WCE1 and WCE2), the standard Gaussian mixture (GM), and the existing robust methods (TCL, IML, CGM), based on the 1000 simulated datasets with $p=5$.
\label{tab:sim-GM2}
}
\begin{center}
{\footnotesize
\begin{tabular}{cccccccccccc}
\hline
& MSE & MSE & MSE &&&MSE & MSE & MSE \\
& $\bmu$ & $\bSig$ & $\bpi$ &MISE& MCE & $\bmu$ & $\bSig$ & $\bpi$ &MISE& MCE   \\
& ($\times 10^2$) & &($\times 10^3$)& ($\times 10^6$) & ($\times 10^2$) & ($\times 10^2$) & &($\times 10^3$)& ($\times 10^6$) & ($\times 10^2$) \\
\hline 
& \multicolumn{5}{c}{$(C=2, \ \omega=0)$} & \multicolumn{5}{c}{$(C=3, \ \omega=0)$}   \\
WCE1 & 1.66 & 2.93 & 2.41 & 1.80 & 6.00 & 1.19 & 2.92 & 1.43 & 1.74 & 2.53 \\
WCE2 & 1.74 & 2.82 & 2.46 & 1.79 & 6.01 & 1.33 & 2.77 & 1.47 & 1.72 & 2.49 \\
GM & 2.23 & 3.58 & 4.19 & 1.97 & 4.90 & 1.07 & 3.70 & 1.34 & 1.93 & 0.32 \\
TCL & 1.64 & 1.31 & 2.27 & 2.21 & 8.98 & 1.21 & 1.01 & 1.58 & 1.69 & 5.14 \\
IML & 2.46 & 1.38 & 3.49 & 1.23 & 5.31 & 1.07 & 0.77 & 1.39 & 0.94 & 0.81 \\
CGM & 1.42 & 1.00 & 1.87 & 1.21 & 4.19 & 1.06 & 0.79 & 1.34 & 1.07 & 0.26 \\
\hline
 & \multicolumn{5}{c}{$(C=2, \ \omega=0.03)$}  & \multicolumn{5}{c}{$(C=3, \ \omega=0.03)$}   \\
WCE1 & 1.77 & 3.10 & 2.88 & 1.73 & 5.91 & 1.23 & 2.91 & 1.50 & 1.64 & 2.24 \\
WCE2 & 1.78 & 2.91 & 2.54 & 1.76 & 5.91 & 1.37 & 2.82 & 1.55 & 1.69 & 2.35 \\
GM & 15.22 & 26.80 & 40.38 & 2.82 & 12.64 & 2.18 & 10.19 & 1.52 & 2.11 & 0.44 \\
TCL & 1.65 & 1.20 & 2.35 & 1.57 & 6.30 & 1.15 & 0.85 & 1.47 & 1.14 & 2.26 \\
IML & 3.70 & 3.02 & 5.98 & 1.35 & 6.00 & 1.12 & 0.87 & 1.42 & 0.99 & 0.96 \\
CGM & 5.61 & 13.12 & 9.81 & 1.47 & 6.20 & 1.18 & 1.68 & 1.46 & 0.98 & 0.28 \\
\hline
 & \multicolumn{5}{c}{$(C=2, \ \omega=0.06)$}  & \multicolumn{5}{c}{$(C=3, \ \omega=0.06)$}  \\
WCE1 & 2.18 & 4.02 & 4.59 & 1.70 & 6.05 & 1.27 & 2.96 & 1.54 & 1.56 & 1.93 \\
WCE2 & 1.88 & 3.09 & 2.98 & 1.74 & 5.89 & 1.39 & 2.84 & 1.60 & 1.63 & 2.13 \\
GM & 30.16 & 59.38 & 66.20 & 3.69 & 17.15 & 5.18 & 32.89 & 2.24 & 3.26 & 0.57 \\
TCL & 2.12 & 2.91 & 2.88 & 1.31 & 4.87 & 1.30 & 1.63 & 1.46 & 0.97 & 0.29 \\
IML & 3.01 & 2.23 & 3.51 & 1.30 & 5.69 & 1.17 & 1.01 & 1.52 & 1.05 & 1.00 \\
CGM & 15.61 & 62.26 & 33.93 & 2.51 & 10.26 & 1.45 & 3.96 & 1.83 & 1.23 & 0.28 \\
\hline
\end{tabular}
}
\end{center}
\end{table}

\begin{table}[!htbp]
\caption{Mean squared errors (MSE) of the model parameters, mean integrated squared errors (MISE) of the density estimation, and mean classification error (MCE) of the proposed methods (WCE1 and WCE2), the standard Gaussian mixture (GM), and the existing robust methods (TCL, IML, CGM), based on the 1000 simulated datasets with $p=10$.
\label{tab:sim-GM3}
}
\begin{center}
{\footnotesize
\begin{tabular}{cccccccccccc}
\hline
& MSE & MSE & MSE &&&MSE & MSE & MSE \\
& $\bmu$ & $\bSig$ & $\bpi$ &MISE& MCE & $\bmu$ & $\bSig$ & $\bpi$ &MISE& MCE   \\
& ($\times 10^2$) & &($\times 10^3$)& ($\times 10^6$) & ($\times 10^2$) & ($\times 10^2$) & &($\times 10^3$)& ($\times 10^6$) & ($\times 10^2$) \\
\hline 
 &  \multicolumn{5}{c}{$(C=2, \ \omega=0)$}  & \multicolumn{5}{c}{$(C=3, \ \omega=0)$}   \\
WCE1 & 0.86 & 6.51 & 25.39 & 8.52 & 10.40 & 0.24 & 4.46 & 1.63 & 7.06 & 2.99 \\
WCE2 & 0.60 & 5.73 & 15.32 & 9.80 & 8.90 & 0.29 & 4.52 & 1.75 & 7.63 & 3.08 \\
GM & 1.29 & 8.26 & 43.01 & 8.60 & 11.56 & 0.20 & 5.29 & 1.39 & 7.42 & 0.37 \\
TCL & 1.99 & 11.86 & 24.72 & 19.24 & 19.99 & 0.22 & 2.66 & 1.60 & 11.18 & 5.23 \\
IML & 2.04 & 10.79 & 20.15 & 9.45 & 14.06 & 0.20 & 2.36 & 1.39 & 5.95 & 0.55 \\
CGM & 0.44 & 3.88 & 7.51 & 7.50 & 5.96 & 0.20 & 2.34 & 1.38 & 5.94 & 0.30 \\
\hline
 & \multicolumn{5}{c}{$(C=2, \ \omega=0.03)$} & \multicolumn{5}{c}{$(C=3, \ \omega=0.03)$} \\
WCE1 & 0.83 & 6.84 & 26.74 & 8.80 & 10.52 & 0.25 & 4.56 & 1.68 & 7.94 & 2.89 \\
WCE2 & 0.60 & 5.89 & 14.83 & 9.62 & 8.95 & 0.30 & 4.65 & 1.83 & 8.58 & 2.98 \\
GM & 3.10 & 36.02 & 95.08 & 10.81 & 21.60 & 0.39 & 15.96 & 1.94 & 9.32 & 0.73 \\
TCL & 2.02 & 12.64 & 26.26 & 16.46 & 17.76 & 0.22 & 2.57 & 1.53 & 8.34 & 2.34 \\
IML & 2.01 & 14.29 & 19.64 & 10.49 & 13.91 & 0.22 & 2.61 & 1.50 & 6.63 & 0.61 \\
CGM & 1.13 & 23.54 & 25.64 & 10.54 & 10.01 & 0.23 & 5.87 & 1.54 & 6.86 & 0.34 \\
\hline
 & \multicolumn{5}{c}{$(C=2, \ \omega=0.06)$}  & \multicolumn{5}{c}{$(C=3, \ \omega=0.06)$}  \\
WCE1 & 1.49 & 11.88 & 65.37 & 11.79 & 15.68 & 0.26 & 4.64 & 1.72 & 7.85 & 2.81 \\
WCE2 & 0.53 & 5.91 & 14.05 & 17.53 & 8.43 & 0.31 & 4.75 & 1.87 & 7.70 & 2.88 \\
GM & 3.54 & 55.76 & 92.74 & 7.92 & 19.89 & 0.87 & 37.46 & 4.88 & 11.25 & 1.48 \\
TCL & 3.12 & 26.48 & 42.54 & 13.23 & 20.68 & 0.25 & 5.05 & 1.43 & 6.90 & 0.45 \\
IML & 1.25 & 10.40 & 21.28 & 11.08 & 11.36 & 0.23 & 2.84 & 1.66 & 6.89 & 0.70 \\
CGM & 1.59 & 34.55 & 43.64 & 9.75 & 12.72 & 0.34 & 15.13 & 2.17 & 7.80 & 0.44 \\

\hline
\end{tabular}
}
\end{center}
\end{table}

\begin{table}[!htbp]
\caption{Percentage of false discovery rate (FDR) and power (PW) of the outlier detection methods under $p\in \{2, 5, 10\}$, averaged over the 1000 simulated datasets.
\label{tab:sim-GM-out}
}
\begin{center}
\begin{tabular}{cccccccccccc}
\hline
& \multicolumn{2}{c}{$p=2$} & \multicolumn{2}{c}{$p=5$} & \multicolumn{2}{c}{$p=10$} \\
& FDR & PW & FDR & PW & FDR & PW \\
\hline \multicolumn{7}{c}{($\xi=2, \ \omega=0.03$)}  \\
WCE1 & 13.8 & 89.8 & 31.0 & 93.6 & 40.0 & 93.9 \\
WCE2 & 15.6 & 91.5 & 32.4 & 93.8 & 41.7 & 93.8 \\
TCL & 40.0 & 93.8 & 39.9 & 93.8 & 39.9 & 93.8 \\
IML & 16.4 & 85.4 & 10.0 & 87.5 & 4.7 & 84.1 \\
CGM & 10.5 & 78.5 & 7.2 & 68.0 & 6.6 & 72.1 \\
\hline \multicolumn{7}{c}{($\xi=2, \ \omega=0.06$)}  \\
WCE1 & 6.0 & 81.9 & 16.2 & 96.1 & 21.2 & 96.8 \\
WCE2 & 7.3 & 91.0 & 18.2 & 96.7 & 25.0 & 96.8 \\
TCL & 1.2 & 79.7 & 0.3 & 80.4 & 0.0 & 80.6 \\
IML & 16.4 & 91.5 & 8.0 & 94.9 & 2.4 & 94.3 \\
CGM & 9.9 & 72.3 & 9.4 & 49.0 & 7.3 & 78.5 \\
\hline \multicolumn{7}{c}{($\xi=3, \ \omega=0.03$)}  \\
WCE1 & 18.3 & 89.8 & 38.4 & 93.9 & 44.7 & 93.9 \\
WCE2 & 20.8 & 91.2 & 39.6 & 93.9 & 45.4 & 94.0 \\
TCL & 40.0 & 93.7 & 39.9 & 93.9 & 39.9 & 93.9 \\
IML & 18.3 & 81.7 & 9.6 & 89.5 & 5.8 & 91.9 \\
CGM & 13.8 & 85.0 & 10.5 & 90.8 & 7.2 & 89.2 \\
\hline \multicolumn{7}{c}{($\xi=3, \ \omega=0.06$)}  \\
WCE1 & 8.4 & 86.7 & 20.7 & 96.8 & 27.8 & 96.9 \\
WCE2 & 10.3 & 91.7 & 22.6 & 96.8 & 28.2 & 96.8 \\
TCL & 2.1 & 79.0 & 0.3 & 80.4 & 0.1 & 80.6 \\
IML & 17.3 & 89.7 & 6.3 & 94.7 & 4.2 & 96.1 \\
CGM & 16.0 & 92.1 & 12.0 & 93.4 & 8.4 & 91.0 \\
\hline
\end{tabular}
\end{center}
\end{table}

\begin{table}[!htbp]
\caption{Mean squared errors (MSE) of the model parameters, mean integrated squared errors (MISE) of the density estimation, and mean classification error (MCE) of the proposed methods (WCE1 and WCE2), the standard Gaussian mixture (GM), and the existing robust methods (TCL, IML, CGM), based on the 1000 simulated datasets under the additional scenarios with $p=2$.
\label{tab:sim-GM-add}
}
\begin{center}
\begin{tabular}{cccccccccccc}
\hline
& MSE & MSE & MSE && &MSE & MSE & MSE&&\\
& $\bmu$ & $\bSig$ & $\bpi$ &MISE& MCE & $\bmu$ & $\bSig$ & $\bpi$ &MISE& MCE \\
& & &($\times 10^3$)& ($\times 10^6$) & ($\times 100$)& &($\times 10^3$)& ($\times 10^6$) & ($\times 100$) \\
\hline 
&\multicolumn{5}{c}{ $(\xi=2)$} &\multicolumn{5}{c}{ $(\xi=3)$}  \\
WCE1 & 3.29 & 60.50 & 22.79 & 5.58 & 5.22 & 0.68 & 45.93 & 2.69 & 4.62 & 0.86 \\ 
WCE2 & 1.70 & 41.50 & 12.05 & 4.50 & 4.79 & 0.26 & 23.61 & 1.88 & 3.63 & 0.94 \\
GM & 9.90 & 82.05 & 66.44 & 22.17 & 13.33 & 3.63 & 83.48 & 6.62 & 13.41 & 0.80\\
TCL & 0.29 & 3.58 & 1.78 & 4.03 & 3.57 & 0.16 & 2.10 & 1.24 & 3.07 & 0.35\\
IML & 15.78 & 8.47 & 110.44 & 40.32 & 27.60 & 57.51 & 8.45 & 194.53 & 59.72 & 40.14\\
CGM & 9.09 & 236.34 & 54.47 & 19.76 & 11.30 & 5.04 & 311.11 & 10.26 & 5.62 & 1.25  \\
\hline
\end{tabular}
\end{center}
\end{table}

\begin{table}[!htbp]
\caption{Proportions ($\%$) of the selected number of components based on the six methods under $p=2$ and $n=500$.
The true number of components ($K$) is $3$.
\label{tab:sim-K}
}
\begin{center}
{\small
\begin{tabular}{ccccccccccccccc}
\hline
&  \multicolumn{5}{c}{$K$} & &  \multicolumn{5}{c}{$K$}\\
 & 2 & 3 & 4 & 5 & 6 &  & 2 & 3 & 4 & 5 & 6 \\
 \hline
 &  \multicolumn{5}{c}{$(C=2, \ \omega=0)$}  &  & \multicolumn{5}{c}{$(C=3, \ \omega=0)$}   \\
WCE1 & 0.0 & 92.3 & 7.7 & 0.0 & 0.0 &  & 0.0 & 97.0 & 3.0 & 0.0 & 0.0 \\
WCE2 & 0.0 & 92.3 & 7.7 & 0.0 & 0.0 &  & 0.0 & 96.0 & 4.0 & 0.0 & 0.0 \\
GM & 0.0 & 100.0 & 0.0 & 0.0 & 0.0 &  & 0.0 & 97.7 & 2.3 & 0.0 & 0.0 \\
TCL & 0.0 & 100.0 & 0.0 & 0.0 & 0.0 &  & 0.0 & 100.0 & 0.0 & 0.0 & 0.0 \\
IML & 23.1 & 53.8 & 23.1 & 0.0 & 0.0 &  & 68.0 & 25.3 & 3.0 & 3.0 & 0.7 \\
CGM & 0.0 & 84.6 & 15.4 & 0.0 & 0.0 &  & 0.0 & 71.0 & 20.7 & 5.0 & 3.3 \\
\hline
 & \multicolumn{5}{c}{$(C=2, \ \omega=0.03)$}  &  & \multicolumn{5}{c}{$(C=3, \ \omega=0.03)$}   \\
WCE1 & 0.0 & 98.7 & 1.3 & 0.0 & 0.0 &  & 0.0 & 98.3 & 1.7 & 0.0 & 0.0 \\
WCE2 & 0.0 & 98.7 & 1.3 & 0.0 & 0.0 &  & 0.0 & 94.9 & 5.1 & 0.0 & 0.0 \\
GM & 0.3 & 92.7 & 7.0 & 0.0 & 0.0 &  & 0.0 & 97.5 & 2.5 & 0.0 & 0.0 \\
TCL & 0.0 & 100.0 & 0.0 & 0.0 & 0.0 &  & 0.0 & 100.0 & 0.0 & 0.0 & 0.0 \\
IML & 14.3 & 63.3 & 15.7 & 4.0 & 2.7 &  & 61.9 & 23.7 & 7.6 & 5.1 & 1.7 \\
CGM & 8.3 & 78.0 & 7.3 & 4.3 & 2.0 &  & 0.0 & 68.6 & 19.5 & 8.5 & 3.4 \\
\hline
 & \multicolumn{5}{c}{$(C=2, \ \omega=0.06)$}   &  & \multicolumn{5}{c}{$(C=3, \ \omega=0.06)$}  \\
WCE1 & 2.0 & 93.3 & 4.3 & 0.3 & 0.0 &  & 0.0 & 97.0 & 3.0 & 0.0 & 0.0 \\
WCE2 & 0.0 & 93.7 & 6.3 & 0.0 & 0.0 &  & 0.0 & 96.0 & 4.0 & 0.0 & 0.0 \\
WCE1 ($\alpha=0.005$) & 3.0 & 92.7 & 3.7 & 0.3 & 0.3 &  & 0.0 & 99.7 & 0.3 & 0.0 & 0.0 \\
WCE2 ($\alpha=0.005$) & 0.0 & 96.7 & 3.3 & 0.0 & 0.0 &  & 0.0 & 99.0 & 1.0 & 0.0 & 0.0 \\
WCE1 ($\alpha=0.001$) & 4.7 & 86.3 & 7.3 & 1.3 & 0.3 &  & 0.0 & 98.0 & 2.0 & 0.0 & 0.0 \\
WCE2 ($\alpha=0.001$) & 1.0 & 93.0 & 5.7 & 0.3 & 0.0 &  & 0.0 & 95.3 & 4.7 & 0.0 & 0.0 \\
GM & 14.0 & 66.3 & 18.7 & 1.0 & 0.0 &  & 0.0 & 81.7 & 17.3 & 1.0 & 0.0 \\
TCL & 0.0 & 96.0 & 4.0 & 0.0 & 0.0 &  & 0.0 & 96.0 & 4.0 & 0.0 & 0.0 \\
IML & 8.0 & 59.7 & 21.0 & 8.7 & 2.7 &  & 56.7 & 31.0 & 6.3 & 4.0 & 2.0 \\
CGM & 43.3 & 33.0 & 14.7 & 7.0 & 2.0 &  & 2.7 & 74.7 & 16.0 & 4.7 & 2.0 \\
\hline
\end{tabular}
}
\end{center}
\end{table}

\subsection{Skew normal mixture}
The performance of the proposed methods for the skew normal mixture is examined using simulated data along with some existing methods.
We consider a $p$-dimensional mixture of skew-normal distributions with $K=2$ components where the mixture components have the location parameter vectors $\bmu_1=-\bmu_2=(\xi,\xi,0,\ldots,0)$, skewness parameter vectors $\bpsi_1=-\bpsi_2=(2,2,0,\ldots,0)$ and scale matrices, $\bSig_1$ and $\bSig_2$ given by $\bSig_1=\bSig_2={\rm blockdiag}(\bSig^{\ast},I_{p-2})$ with $(\bSig^{\ast})_{11}=(\bSig^{\ast})_{22}=1$ and $(\bSig^{\ast})_{12}=(\bSig^{\ast})_{21}=0.3$, in the parametrization given in \eqref{Hie}.
The two cases for $\xi$, $\xi=1$, and $\xi=2$ are considered, corresponding to the cases of the overlapping and well-separated clusters, respectively.
The mixing proportions are set to $\pi_1=0.4$ and $\pi_2=0.6$.
The sample size is set to $n=500$, and the results for $n=2000$ are provided in the Supplementary Material. 
The proportion of outliers is set to $\omega\in \{0.03, 0.06, 0.09\}$, and the outliers are generated in the same way as in Section \ref{sec:sim-GM}.

\begin{figure}[!htb]
\centering
\includegraphics[width=13cm,clip]{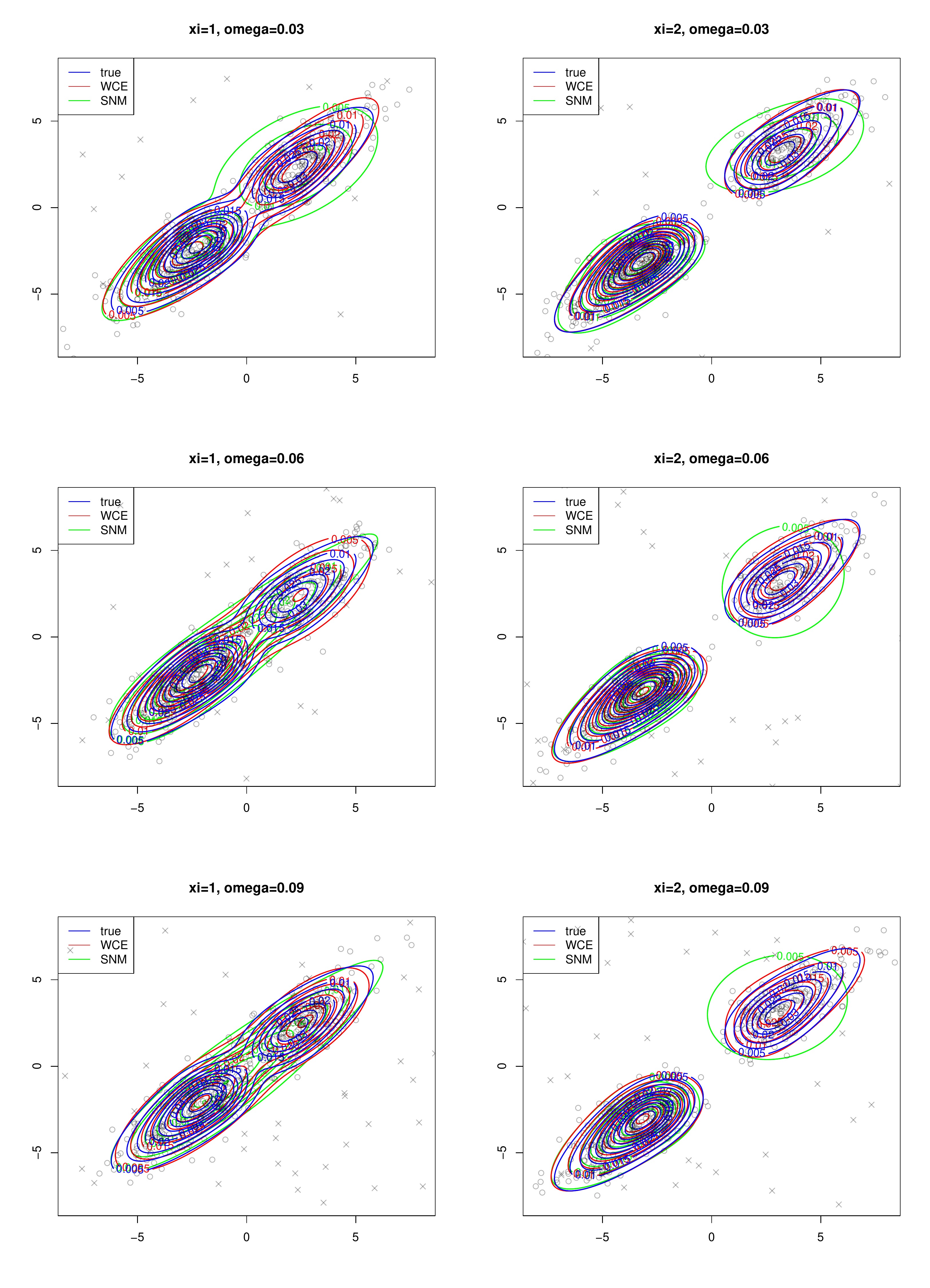}
\caption{The contour plots of the true density function (blue), and estimated density functions based on WCE with $\gamma=0.2$ (red) and the standard maximum likelihood method (green) applied to simulated datasets with the 6 combinations of $\xi\in \{1, 2\}$ and $\omega\in \{0.03, 0.06, 0.09\}$. 
\label{fig:one-shot}
}
\end{figure}

First, we show some results based on a single simulated dataset with $p=2$. 
The proposed weighted complete equation (WCE) method is applied to fit the skew normal mixture with $\gamma=0.2$. 
The standard skew normal mixture (SNM) is estimated using the maximum likelihood method with the EM algorithm of \cite{Lin2007}.
Figure \ref{fig:one-shot} presents the three contours of the estimated and true densities.
The figure shows that the proposed WCE method provides reasonable estimates of the true density by adequately suppressing the influence of outliers through density weights.
On the other hand, SMN produces inaccurate estimates. 
In particular, it overestimates the variability of the true data-generating process by treating the outliers as non-outliers. 
It is also observed that the inaccuracy of the SNM becomes more profound as the proportion of outliers increases.

Next, the performance of the proposed method and the existing methods are quantitatively compared. 
The two settings for the proposed WCE method with $\gamma=0.2$ and $\gamma=0.3$ are considered, denoted as WCE1 and WCE2, respectively.
As contenders, we consider SNM and the skew-$t$ mixture \citep{Lee2016} of the R package `EMMIXcskew' \citep{Lee2017} denoted by STM. 
Note that STM is also robust against outliers because of the heavy tails of the skew-$t$ distribution, but the interpretations of the skewness parameters and scale matrices are different from those for WCE and SNM.
Thus, the comparison with STM is made only with respect to the mean (location) parameters, mixing proportions, and classification accuracy. 
We considered two cases of $p$, $p=2$ and $p=5$, but the computing time for STM under $p=5$ is too long to include STM as a contender in our study, so the results of STM are reported only for $p=2$.
To evaluate this estimation performance, we employ the squared errors: $\sum_{k\in \{1,2\}}\|\widehat{\bmu}_k-\bmu_k\|^2$, $\sum_{k\in \{1,2\}}\|\widehat{\bpsi}_k-\bpsi_k\|^2$, $\sum_{k\in \{1,2\}}\|{\rm vec}(\widehat{\bSig}_k)-{\rm vec}(\bSig_k)\|^2$ and $\sum_{k\in \{1,2\}}(\widehat{\pi}_k-\pi_k)^2$.
Based on 1000 replications, the mean squared error (MSE) for the model parameters is obtained. 
We also computed the mean integrated squared error (MISE) of the density estimation \eqref{MISE} and the mean classification error in the same way as in Section \ref{sec:sim-GM}.  
The table shows that the standard SNM model is severely affected by the outliers resulting in a large MSE, as in the case of the normal mixture.
In the present setting, MCE for SNM in the case of overlapped clusters is particularly large.
Comparing the proposed method and STM for $p=2$, it can be seen that the proposed method resulted in a smaller MSE for $\bmu$ and $\bpi$ than STM, while MCE appears somewhat comparable.
The table shows that the proposed method also seems to work reasonably well in the skew normal mixture case and suggests that our method is a useful tool applicable to a wide range of mixture models.

\begin{table}[!htbp]
\caption{Performance measures for the two proposed methods (WCE1 and WCE2), standard skew normal (SNM) mixture and skew-$t$ mixture (STM) under $p=2$ and $n=500$. 
The MSE values for $\bpi$ are multiplied by 100. 
\label{tab:sim-SNM1}
}
\begin{center}
\begin{tabular}{ccccccccccccccc}
\hline
$p=2$& \multicolumn{4}{c}{MSE} & MISE& MCE\\
& $\bmu$ & $\bSig$ & $\bpsi$ &$\bpi$ & ($\times 10^5$) & (\%)\\
\hline 
&\multicolumn{6}{c}{($\xi=1, \ \omega=0.03$)} \\
WCE1 & 1.22 & 8.59 & 2.15 & 0.48 & 0.66 & 2.94 \\
WCE2 & 1.43 & 10.97 & 2.76 & 0.43 & 0.61 & 2.83 \\
SNM & 25.48 & 29.00 & 38.70 & 21.61 & 2.42 & 24.47 \\
STM & 5.62 & - & - & 1.93 & - & 4.12 \\
\hline 
&\multicolumn{6}{c}{($\xi=1, \ \omega=0.06$)} \\
WCE1 & 1.29 & 10.64 & 2.33 & 0.38 & 0.72 & 3.66 \\
WCE2 & 1.70 & 12.75 & 3.30 & 0.32 & 0.57 & 3.62 \\
SNM & 46.21 & 61.28 & 75.54 & 39.67 & 3.22 & 47.80 \\
STM & 7.87 & - & - & 3.88 & - & 7.02 \\
\hline
&\multicolumn{6}{c}{($\xi=1, \ \omega=0.09$)} \\
WCE1 & 2.11 & 16.40 & 4.54 & 1.60 & 1.06 & 6.36 \\
WCE2 & 2.17 & 15.08 & 4.10 & 0.48 & 0.63 & 4.51 \\
SNM & 45.16 & 80.68 & 76.50 & 36.66 & 3.62 & 49.63 \\
STM & 11.20 & - & - & 7.82 & - & 14.83 \\
\hline 
&\multicolumn{6}{c}{($\xi=2, \ \omega=0.03$)} \\
WCE1 & 0.69 & 2.14 & 1.21 & 0.10 & 0.31 & 0.66 \\
WCE2 & 0.71 & 2.47 & 1.33 & 0.10 & 0.32 & 0.65 \\
SNM & 1.19 & 8.98 & 2.28 & 0.12 & 2.11 & 0.18 \\
STM & 5.95 & - & - & 0.87 & - & 0.91 \\
\hline 
&\multicolumn{6}{c}{($\xi=2, \ \omega=0.06$)} \\
WCE1 & 0.69 & 2.48 & 1.19 & 0.09 & 0.40 & 1.40 \\
WCE2 & 0.86 & 3.22 & 1.59 & 0.09 & 0.38 & 1.32 \\
SNM & 1.25 & 25.51 & 2.85 & 0.22 & 3.28 & 0.33 \\
STM & 7.85 & - & - & 0.89 & - & 1.07 \\
\hline 
&\multicolumn{6}{c}{($\xi=2, \ \omega=0.09$)} \\
WCE1 & 0.86 & 3.24 & 1.47 & 0.11 & 0.58 & 2.28 \\
WCE2 & 1.12 & 4.09 & 2.01 & 0.10 & 0.48 & 2.03 \\
SNM & 1.28 & 44.31 & 3.21 & 0.46 & 3.94 & 0.47 \\
STM & 8.94 & - & - & 1.34 & - & 1.47 \\
\hline
\end{tabular}
\end{center}
\end{table}

\begin{table}[!htbp]
\caption{Performance measures for the two proposed methods (WCE1 and WCE2), standard skew normal (SNM) mixture and skew-$t$ mixture (STM) under $p=5$ and $n=500$. 
The MSE values for $\bpi$ are multiplied by 100. 
\label{tab:sim-SNM2}
}
\begin{center}
\begin{tabular}{ccccccccccccccc}
\hline
$p=5$ & \multicolumn{4}{c}{MSE} & MISE& MCE\\
& $\bmu$ & $\bSig$ & $\bpsi$ &$\bpi$ & ($\times 10^{10}$) & (\%)\\
\hline 
&\multicolumn{6}{c}{($\xi=1, \ \omega=0.03$)} \\
WCE1 & 1.74 & 0.39 & 4.16 & 0.31 & 0.29 & 2.55 \\
WCE2 & 1.83 & 0.40 & 4.28 & 0.32 & 0.32 & 2.65 \\
SNM & 2.22 & 6.66 & 5.40 & 1.48 & 1.23 & 4.12 \\
\hline 
&\multicolumn{6}{c}{($\xi=1, \ \omega=0.06$)} \\
WCE1 & 0.51 & 0.37 & 1.08 & 0.16 & 0.26 & 2.11 \\
WCE2 & 0.56 & 0.38 & 1.11 & 0.16 & 0.28 & 2.10 \\
SNM & 1.08 & 20.72 & 2.57 & 0.81 & 1.85 & 3.02 \\
\hline 
&\multicolumn{6}{c}{($\xi=1, \ \omega=0.09$)} \\
WCE1 & 0.28 & 0.35 & 0.46 & 0.13 & 0.29 & 2.24 \\
WCE2 & 0.31 & 0.37 & 0.46 & 0.13 & 0.29 & 2.23 \\
SNM & 1.72 & 36.35 & 3.49 & 1.20 & 2.15 & 3.07 \\
\hline 
&\multicolumn{6}{c}{($\xi=2, \ \omega=0.03$)} \\
WCE1 & 0.30 & 0.38 & 0.39 & 0.09 & 0.27 & 0.17 \\
WCE2 & 0.34 & 0.38 & 0.44 & 0.10 & 0.28 & 0.17 \\
SNM & 0.44 & 6.51 & 0.80 & 0.11 & 1.15 & 0.19 \\
\hline 
&\multicolumn{6}{c}{($\xi=2, \ \omega=0.06$)} \\
WCE1 & 0.26 & 0.36 & 0.36 & 0.11 & 0.26 & 0.36 \\
WCE2 & 0.29 & 0.38 & 0.38 & 0.11 & 0.28 & 0.35 \\
SNM & 1.52 & 23.78 & 3.24 & 0.25 & 1.93 & 0.38 \\
\hline 
&\multicolumn{6}{c}{($\xi=2, \ \omega=0.09$)} \\
WCE1 & 0.23 & 0.35 & 0.34 & 0.10 & 0.28 & 0.58 \\
WCE2 & 0.30 & 0.37 & 0.41 & 0.10 & 0.29 & 0.55 \\
SNM & 3.06 & 40.12 & 6.80 & 0.52 & 2.21 & 0.53 \\
\hline
\end{tabular}
\end{center}
\end{table}

\section{Real data example}\label{sec:app}

\subsection{Tone perception data}\label{sec:tone}
\begin{figure}[!htb]
\centering
\includegraphics[width=13cm,clip]{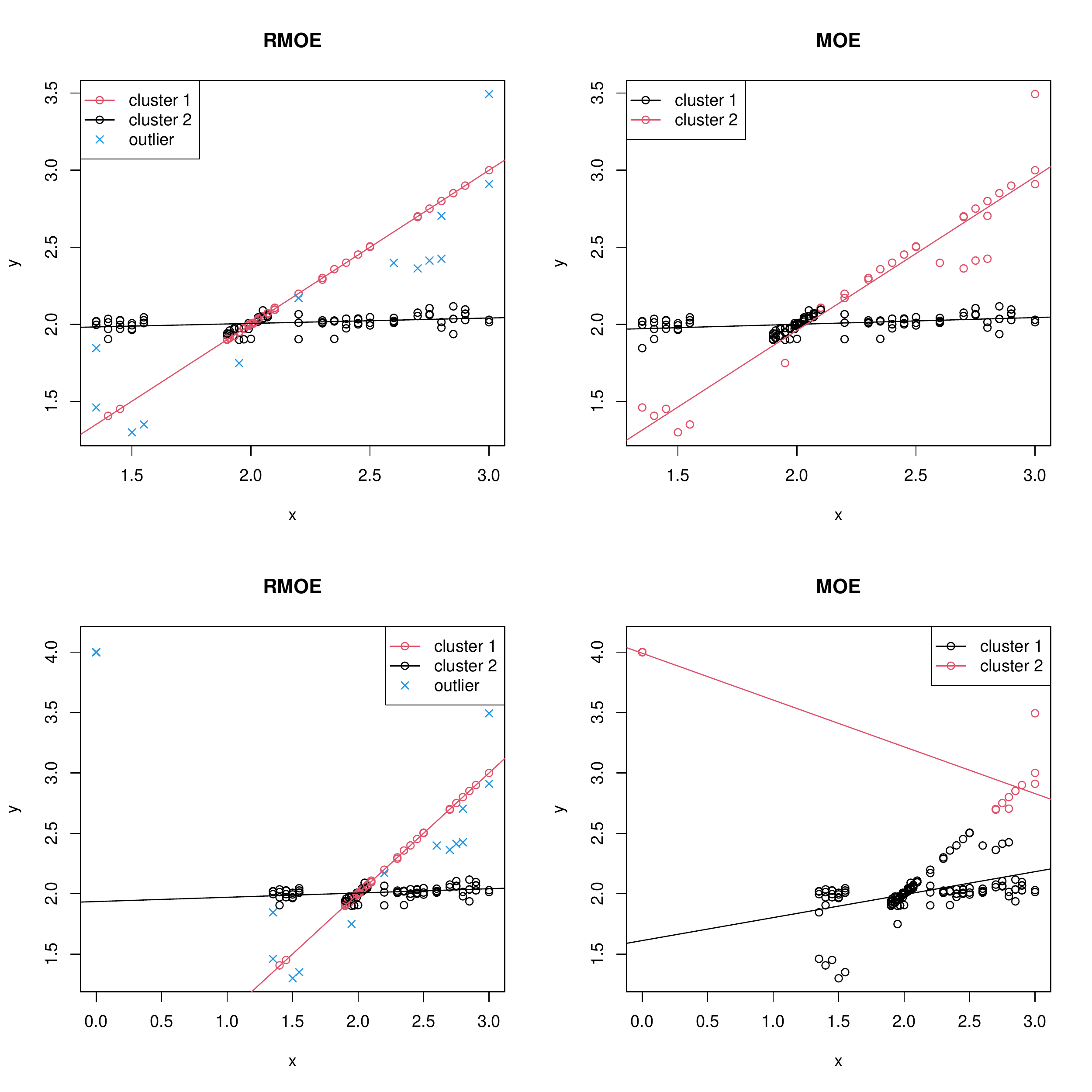}
\caption{Estimated regression lines in the two clusters, detected outliers and classification results for the tone perception data based on the standard maximum likelihood (MOE) and robust (RMOE) method for the original tone data (upper) and contaminated data (lower). 
\label{fig:tone}
}
\end{figure}

The proposed robust mixture of experts modeling is applied to a famous tone perception dataset \citep{Cohen1984}. 
In the tone perception experiment, a pure fundamental tone added with electronically generated overtones determined by a stretching ratio was played to a trained musician. 
The data consists of $n=150$ pairs of ``stretch ratio" variable (denoted by $Y$) and ``tuned" variable (denoted by $X$) and the conditional distribution of $Y$ given $X=x$ is of interest. 
Following \cite{CH2016}, we consider the two-component mixture of experts model given by
\begin{align*}
f(y_i|x_i)&=g(\bx_i;\bEta)\phi(y_i; \beta_{01}+\beta_{11}x_i,\sigma_1^2)\\
&+\{1-g(x_i;\bEta)\}\phi(y_i;\beta_{02}+\beta_{12}x_i, \sigma_2^2),
\end{align*}
where $g(x_i;\bEta)={\rm logit}^{-1}(\eta_0+\eta_1x_i)$. 
The unknown parameters are estimated based on two approaches: the standard maximum likelihood method via the EM algorithm denoted by MOE and the proposed WCE method with $\gamma=0.3$ denoted by RMOE.
Based on the outlier detection rule in Section \ref{sec:outlier}, we detected 13 outliers for $\alpha\in\{0.001, 0.005, 0.01\}$.  
The estimated regression lines, detected outliers, and classification results are shown in the upper panels of Figure \ref{fig:tone}. 
The figure shows that the estimated regression lines based on the two methods are slightly different, but both methods successfully capture the grouped structure of the dataset. 
It is also found that the estimate of $\eta_1$ is almost equal to $0$, thus the mixing probability is homogeneous.

Next, we investigated the sensitivity of the model against outliers by artificially adding the 10 identical pairs $(0,4)$ to the original dataset as outliers, as done in \cite{CH2016}.   
The estimated regression lines obtained from the two methods are shown in the lower panels of Figure \ref{fig:tone}.
The figure clearly shows that MOE is very sensitive to outliers. 
In contrast, the estimates of the proposed RMOE for the original and contaminated datasets are quite similar, indicating the desirable robustness properties of the proposed method. 
Different values of $\gamma\in \{0.3, 0.4, 0.5\}$ are also tested, but the results are not sensitive to $\gamma$. 
We also note that the result of the proposed method in Figure \ref{fig:tone} is almost identical to that reported by \cite{CH2016}, where the $t$-distribution was used as a robust alternative. 
Finally, using the outlier detection rule, we identified 23 outliers, including 10 artificial outliers for any $\alpha\in\{0.001, 0.005, 0.01\}$.

\subsection{Swiss banknote data}\label{sec:swiss}
The proposed WCE method is applied to the famous Swiss banknote dataset \citep{FR1988}.
The data consists of six measurements ($p=6$) made on the 100 genuine and 100 counterfeit old Swiss 1000 franc bills.
We apply the skew normal mixture model to the data using the proposed WCE and the standard EM algorithm.

\begin{figure}[!htb]
\centering
\includegraphics[width=8.5cm,clip]{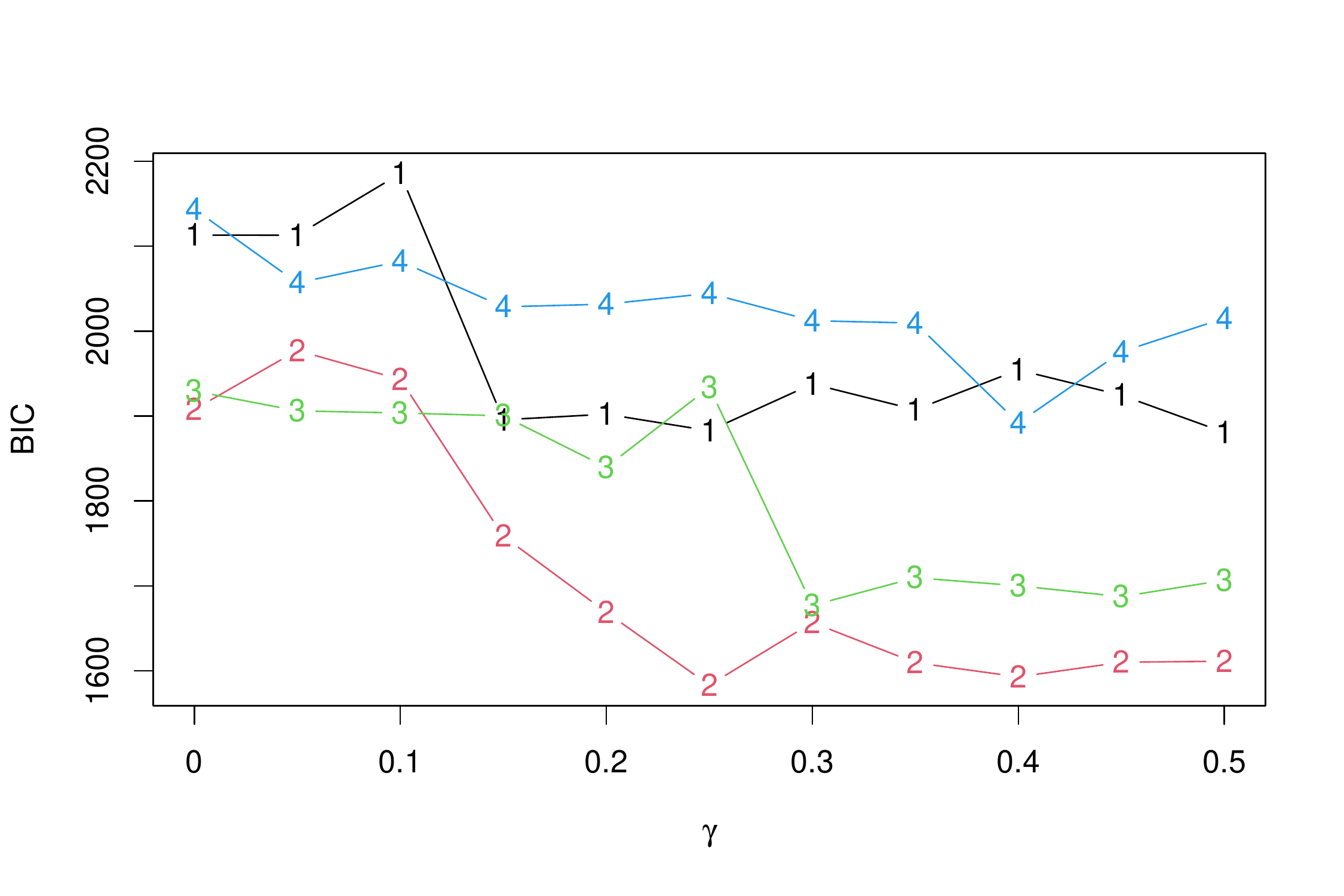}
\caption{Monitoring the trimmed BIC criterion for $K\in \{1,2,3,4\}$ and $\gamma$ under the skew normal mixture models applied to Swiss banknote data. 
\label{fig:bank-BIC}
}
\end{figure}

We first select the number of clusters $K$. 
To this end, we computed the trimmed BIC criterion (\ref{SBIC}) for $K\in \{1,2,3,4\}$ and $\gamma\in\{0, 0.05, \ldots, 0.45, 0.5\}$ in which we set $\alpha=0.001$ (threshold probability for outliers) and $c=50$ (eigen-ratio constraint). 
Note that we also carried out the same procedure with $\alpha\in\{0.005, 0.01\}$ and $c\in\{20, 30\}$, but the results were almost identical. 
The results shown in Figure \ref{fig:bank-BIC} indicate that $K=2$ is a reasonable choice for the number of clusters, which is consistent with the true number of labels. 
We also note that under the non-robust method ($\gamma=0$), the BIC values of $K=2$ and $K=3$ are almost identical.

\begin{figure}[!htb]
\centering
\includegraphics[width=13cm,clip]{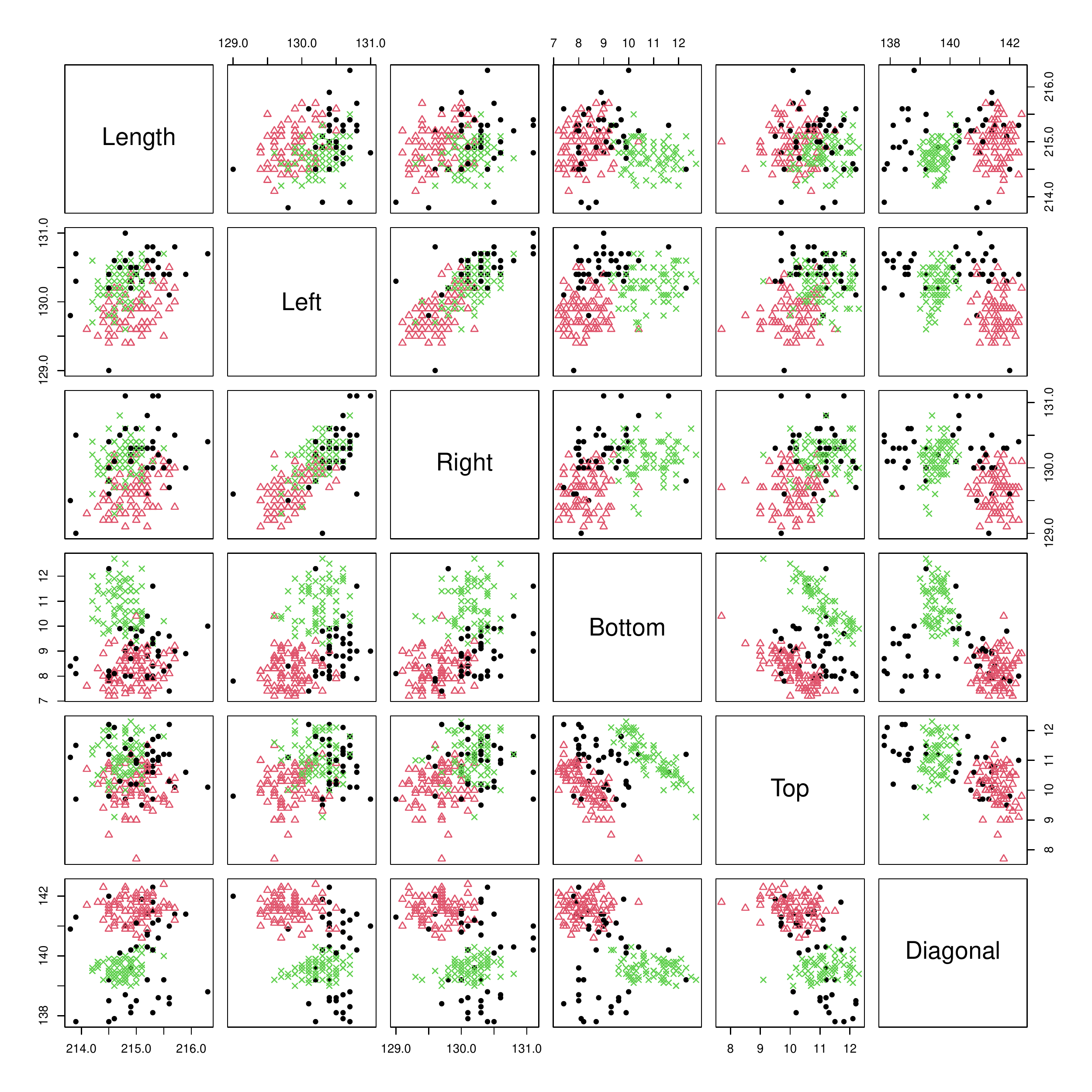}
\caption{Cluster assignment and detected outliers by the proposed WCE method with the skew normal mixture models applied to the bank note data.
The genuine bills are denoted by a green $\times$, counterfeit bills by a red $\triangle$, and outliers by a black filled circle. 
\label{fig:bank}
}
\end{figure}

Using $\gamma=0.3$ in the proposed WCE method, we fitted the skew normal mixture models with $K=2$. 
The robust estimates and standards errors of the skewness parameter $\bpsi_k$ are listed in Table \ref{tab:bank}. 
The table shows significant skewness in the two measurements (bottom and top), so the skew-normal mixture models would be more desirable than Gaussian mixture models for this dataset. 
By carrying out outlier detection with $\alpha=0.001$, we identified 33 outliers, including 14 genuine and 19 counterfeit bills.  
The outliers and resulting cluster assignments are shown in Figure \ref{fig:bank}.
The figure shows that the skew normal mixture can flexibly capture the cluster-wise distributions of the six measurements while successfully suppressing the influence of the outliers.

Finally, we assessed the classification accuracy of the proposed WCE method and standard (non-robust) EM algorithm. 
Based on the fitted results, the cluster assignment for all observations including outliers is obtained as the maximizer of the posterior probability.
The estimated assignments with the true labels ("genuine" or "counterfeit") are then compared. 
The proposed method was able to classify all bills correctly, that is, there was no misclassification, whereas the standard EM algorithm misclassified 18 bills. 
This result is consistent with that of the simulation study and thus strongly suggests using the robust mixture model to correctly classify the observations to the clusters.

\begin{table}[!htbp]
\caption{Robust estimates and standard errors of the skewness parameters in the skew normal mixture models applied to the Swiss banknote data.  
\label{tab:bank}
}
\begin{center}
\begin{tabular}{ccccccccccccccc}
\hline
& \multicolumn{2}{c}{Cluster 1} & \multicolumn{2}{c}{Cluster 2}\\
 & Estimate & SE & Estimate & SE \\
 \hline
Length & -0.02 & 0.17 & -0.11 & 0.55 \\
Left & -0.09 & 0.05 & 0.09 & 0.26 \\
Right & -0.05 & 0.18 & 0.01 & 0.15 \\
Bottom & 0.82 & 0.08 & 1.33 & 0.24 \\
Top & -0.95 & 0.07 & -1.00 & 0.23 \\
Diagonal & 0.14 & 0.11 & -0.05 & 0.16 \\
\hline
\end{tabular}
\end{center}
\end{table}

\section{Concluding remarks}\label{sec:dis}
In this paper, we have introduced a new strategy for robust mixture modeling based on the weighted complete estimating equations.
We then developed an EEE algorithm that iteratively solves the proposed estimating equations. 
The advantage of the proposed method is its computational feasibility because the proposed EEE algorithm admits relatively simple iterations. 
Applications of the proposed method to the three mixture models (multivariate Gaussian mixture, mixture of experts, and  multivariate skew normal mixture) were considered, and feasible EEE algorithms to estimate these models were derived. 
In particular, for the skew normal mixture, we have slightly extended the proposed method and derived a novel EEE algorithm in which all parameters are updated in closed forms. 
The desirable performance of the proposed method is confirmed through a simulation study and real data examples.

In this work, we only considered the unstructured variance-covariance (scale) matrix of the Gaussian mixture and skew normal mixture in Sections \ref{sec:GM} and \ref{sec:SNM}. 
However, it might be useful to consider a structured scale matrix characterized by some parameters as done in R package `Mclust' \citep{Mclust} or selecting a suitable structure of the scale matrix via an information criterion \citep[e.g.][]{Cerioli2018}.
Furthermore, while $\gamma$ is selected by monitoring the trimmed BIC defined in (\ref{SBIC}) in this paper, it might be possible to employ a selection criterion for robust divergence, suggested by, for example, \citep{basak2021optimal} and \cite{sugasawa2021selection}.
Incorporating such approaches into the proposed method will be a valuable future study.

As briefly mentioned in Section \ref{sec:intro}, the proposed weighted complete estimating equations are closely related to the density power divergence \citep{Basu1998}, since the derivative of the density power divergence of the component-wise distribution given the latent membership variables leads to the weighted complete estimating equations for the component-wise parameters.
Hence, other types of divergence can also be adopted to devise alternative estimating equations. 
However, there is no guarantee that a feasible EEE algorithm, as developed in the present study, can be obtained for other divergences.
An investigation of the usage of another divergence and its comparisons exceeds the scope of this study and is thus left for future research.

\section*{Acknowledgements}
We would like to thank the three reviewers and the coordinating editor for their valuable comments and suggestions, which have significantly improved the paper.  
This work was supported by the Japan Society for the Promotion of Science (KAKENHI) Grant Numbers 18K12754, 18K12757, 20H00080, 21K01421, and 21H00699.

\bibliographystyle{chicago}   
\bibliography{refs}  

\end{document}